\documentclass[12pt]{iopart}
\usepackage{citesort}

\usepackage{graphicx}
\usepackage{epstopdf}
\usepackage{subfigure}
\usepackage{booktabs}
\usepackage{multicol}  
\usepackage{multirow} 
\usepackage{tabularx}
\usepackage{enumerate}
\usepackage{color}
\begin{document}

\title{Chinese cities' air quality pattern and correlation}

\author{Wenjun Zhang$^1$, Zhanpeng Guan$^2$, Jianyao Li$^2$, \\ Zhu Su$^{3,*}$, Weibing Deng$^{1,*}$ and Wei Li$^{1,*}$}
\address{$^1$ Key Laboratory of Quark and Lepton Physics (MOE) and Institute of Particle Physics,  Central China Normal University, Wuhan 430079, China}
\address{$^2$ College of Physical Science and Technology,  Central China Normal University, Wuhan 430079, China}
\address{$^3$ National Engineering Laboratory for Educational Big Data, Central China Normal University, Wuhan 430079, China}

\ead{suz@mail.ccnu.edu.cn, wdeng@mail.ccnu.edu.cn, liw@mail.ccnu.edu.cn}

\vspace{10pt}
\begin{indented}
\item[]\today
\end{indented}

\begin{abstract}
Air quality impacts people's health and daily life, affects the sensitive ecosystems, and even restrains a country's development. By collecting and processing the time series data of Air Quality Index (AQI) of 363 cities of China from Jan. 2015 to Mar. 2019, we dedicated to characterize the universal patterns, the clustering and correlation of air quality of different cities by using the methods of complex network and time series analysis. The main results are as follows: 1) The Air Quality Network of China (AQNC) is constructed by using the Planar Maximally Filtered Graph (PMFG) method. The geographical distances on the correlation of air quality of different cities have been studied, it is found that 100 km is a critical distance for strong correlation. 2) Eight communities of AQNC have been detected, and their patterns have been analyzed by taking into account the Hurst exponent and climate environment, it is shown that the eight communities are reasonable, and they are significantly influenced by the climate factors, such as monsoon, precipitation, geographical regions, etc. 3) The motifs of air quality time series of eight communities have been investigated by the visibility graph, for some communities, the evolutionary patterns of the motifs are a bit stable, and they have the long-term memory effects. While for others, there are no stable patterns.

\end{abstract}

%
\noindent{\textbf{Keywords}}: Air quality, network, community, Hurst exponent, motif, visibility graph

%
%
%

\section{Introduction}

Air pollution has attracted increasing attention in recent years, due to its negative effects on the human health and environmental problems \cite{kenneth1998air,Pope1995Particulate,fakinle2016air}. The relevant interesting questions include the spatial-temporal pattern of air quality, the propagation of air pollution, the relations between air quality and earth environment, etc.  Previous studies in this domain would be generally divided into two groups: 

1) {\it Time series analysis}: It was often used to describe some basic features  of air quality data \cite{Brockwell1989Time}. For examples, Schwartz \cite{Schwartz1990Mortality} found that there are strong correlations between air pollution levels and daily mortality in London. Kim \cite{kim2017ordinal} proposed a generalized linear model based on the time series data of ozone in Southern California, the model can effectively capture the seasonal non-stationary in ordinary time series. The relation between AQI and social-economical factors was also studied in \cite{Xu2016Spatiotemporal}, they analyzed the AQI of 31 provincial cities of China, and found that the value of AQI is positively correlated with the economic level and population level. Li \cite{li2016time} tried to infer the urban air condition from perspective of time series, they focused on $\rm PM_{2.5}$ based urban air quality, and introduced two kinds of time-series methods for real-time and fine-grained air quality prediction. They also proposed a model to show that the spatial scaling rules of population, roads and socioeconomic interactions are in a consistent framework \cite{li2017simple}. Xu \cite{xu2017clearer} presented a data analysis framework to uncover the impact of urban traffic on estimating air quality in different locations within a metropolitan area. They estimated the commuter's exposure to ambient $\rm PM_{2.5}$ by using the mobile phone data, the environmental justice in $\rm PM_{2.5}$ exposure was investigated by comparing the exposure with housing price \cite{xu2019unraveling}.

2) {\it Complex networks}: It is an active area of studying the non-trivial topological features, and relations within the multi-agent systems \cite{erdHos1960evolution,Watts1998Collective,Barabasi1999Emergence}. It could be also used to study the features and evolution of the time series data \cite{lacasa2008time,zhang2006complex,marwan2009complex,yang2008complex,bezsudnov2014time,gao2017complex}.   Representative works, such as, Fan \cite{fan2016characterizing} studied the PM 2.5 time series data by networks. The phase spaces are denoted as nodes, and edges are assigned to nodes with higher correlation coefficients. They analyzed the relations between the criteria of correlation coefficients and the topological quantities,  the similarities of different cities' air quality. Carnevale \cite{carnevale2009neuro} use neural network to find the source of air pollutants, and found that the source of PM 10 is the easiest and most accurate to be located. Zhang \cite{Zhang2018Correlation} studied the correlation and scaling behaviors of PM 2.5 time series of different cities of China, and found that the probability distribution of the correlations has two peaks, the weighted degree distributions of networks with different kinds of correlations are also discussed. Du \cite{du2019percolation} established the correlation network by using the AQI datasets of 35 major cities, they found that the abrupt phase transition usually occurs between three to six weeks ahead of the peak or valley point of the evolution of the AQIs mean in highly polluted region. Zhang \cite{zhang2018research} studied the AQI datasets of Beijing, they transformed the AQI time series to symbol sequence, and studied the different patterns from the network perspective. Wei \cite{wei2019complex} studied the criticality evaluation of air quality standards by the network approach. the network was constructed by the relations of each standard, the critical standards are identified by measuring the centrality of nodes in the network.

Time series analysis and complex networks are two useful metrics for carrying out the quantitative analysis of the air quality data. However, most of the previous works did not consider different cities as a whole system (other than the fluctuations, there should be interactions and correlations), and the geographical factors are not taken into account when analyzing the temporal characteristic of air quality. Therefore, we dedicate to study both the air quality patterns and the correlations of different cities with more complete datasets as we can \cite{tianqihoubao}, that is, the AQI series data of totally 363 cities from Jan. 2015 to Mar. 2019. 

Through calculating the Pearson Correlation Coefficients \cite{benesty2009pearson} of AQI series data between each pair of cities, we construct the AQNC by using the PMFG algorithm \cite{Zhao2016Structure,Kenett2015Network,tumminello2005tool}. 
The probability distribution of the geographical distances of cities  which have direct links in AQNC shows that the air pollution has a strong correlation within 100 km and this correlation would become weak as the distance increases. Eight communities are detected in AQNC based on the edges centrality algorithm \cite{sun2013methods}, the detection results are reasonable both for the large modularity and the geographical distribution. The air quality patterns of each community are studied by considering the long-term memory effects and geographical environments. To uncover the characteristics of the motifs of AQI series data, we use the visibility graph to explore the evolutionary patterns of motifs.

The rest of the paper is organized as follows. In Section 2, the AQNC is constructed by PMFG, the basic properties of AQI and the correlations of air quality of different cities also studied. In Section 3, the community structure of AQNC is detected, and the patterns of different communities are analyzed by taking into account the regional average Hurst exponents, average Hurst exponents of each city, etc. Section 4 shows the results of AQI evolutionary pattern using the visibility graph method. Conclusions and discussions are made in Section 5.

\section{Construction of AQNC and Correlation of AQI of different cities}

The comprehensive evaluation of AQI is based on the pollution index of $SO_2$, $NO_2$, $CO$, $PM_{2.5}$, $PM_{10}$, and $O_3$ \cite{china2015ministry}. We collected and processed the daily AQI time series data of 363 cities of China from Jan. 2015 to Mar. 2019 \cite{tianqihoubao}. The AQI time series data is transformed to AQNC by employing the well-known PMFG method. Based on the correlation of AQI of different cities, the geographical distance on the air pollution diffusion has been investigated.

\subsection{The construction of AQNC and some basic properties of AQI}

\begin{figure}
\subfigcapskip=5pt
\begin{tabular}{cc}
\subfigure[The  $ AQI$ and trend curve of Akesu.]{\label{fig:detrend:a}
\includegraphics[width=0.45\textwidth]{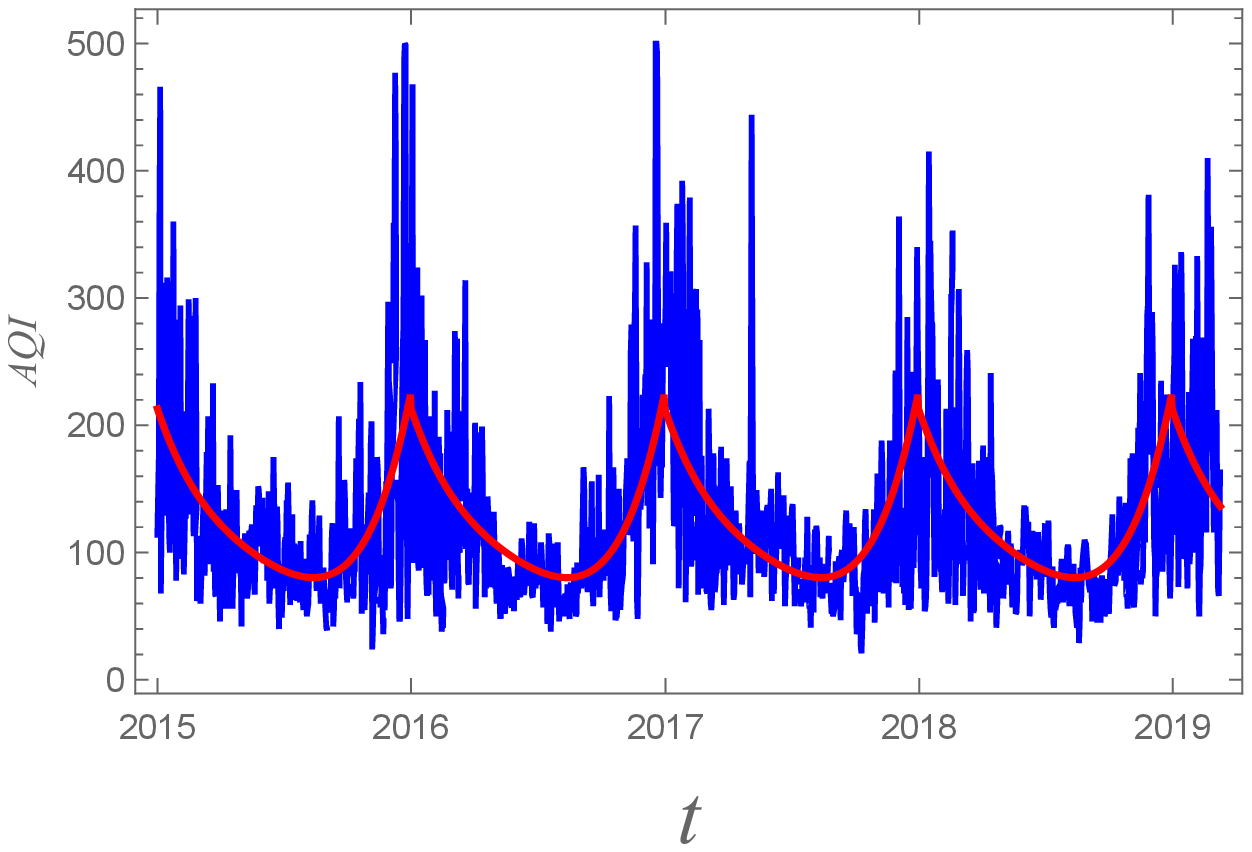}}
&
\subfigure[Time evolution of  $AQI_d$ of Akesu.]{\label{fig:detrend:b}
\includegraphics[width=0.45\textwidth]{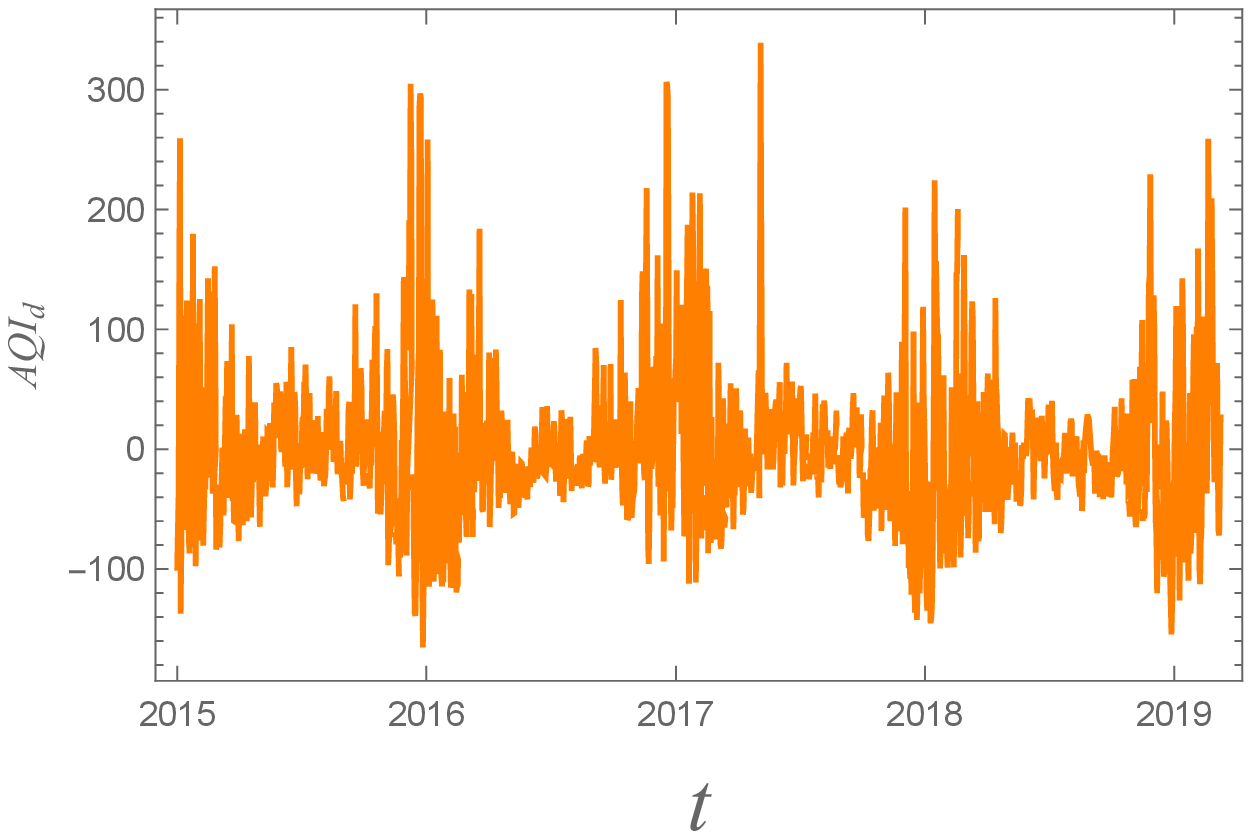}}
\end{tabular}
\caption{(Color online) (a) The $\rm AQI$ time series of Akesu (blue curve), the red line is the fitting curve by employing the polynomial function. (b) The time series of the detrended datasets of AQI (denoted as $\rm AQI_d$).}
\label{fig:AQItrend}
\end{figure}

The air quality displays seasonal cycles, thus the datasets of AQI have strong seasonal effect. There are different methods of removing this effect, such as by subtracting the mean seasonal cycle and dividing by the seasonal standard deviation of each grid point time series \cite{fan2017network,meng2017percolation} or the curve-fitting method \cite{brownlee2017introduction}. The time length of our AQI datasets are around 4 years, the mean seasonal cycle would have large fluctuations, therefore we decide to use the curve-fitting method to remove the effect of seasonality, the calculation processes are as follows:
\begin{enumerate}[1)]
\item Fit the AQI time series data of each year by using the polynomial function with 5 parameters, so as to get the general trend of the datasets. 
\item The detrended datasets (denoted as $\rm AQI_d$) are obtained by subtracting the trend-datasets from the original datasets.
\end{enumerate} 
One example of this detrended process is shown in Fig.~\ref{fig:AQItrend} for the city of Akesu. The blue and red cures on Fig.~\ref{fig:detrend:a} represent the evolution of the original datasets and fitting trend datasets of AQI, respectively. The orange cure on Fig.~\ref{fig:detrend:b} corresponds to the evolution of the detrended datasets of AQI (denoted as  $\rm AQI_d$). We have tested the seasonality trends of the $\rm AQI_d$ by employing the Fourier transform method \cite{bloomfield2004fourier}, and found that the periodogram has no obvious spike. We also calculate the cycle lengths of the top 10 peaks, all of them are not close to one year. Therefore, we would conclude that the seasonality trends of the detrended AQI data ($\rm AQI_d$) are removed.

The Pearson Correlation Coefficient is introduced to calculate the correlations of the detrended AQI time series data $\rm AQI_d$ of $N=363$ cities. The $C_{ij}^{t,\delta}$ between cities $i$ and $j$ during the time period $\{t-\delta/2,t+\delta/2\}$ is defined as
\begin{equation}
C_{ij}^{t,\delta}=\frac{\langle I_i^tI_j^t\rangle-\langle I_i^t\rangle\langle I_j^t\rangle}{\sqrt{[\langle{I_i^t}^2\rangle-{\langle I_i^t\rangle}^2][\langle{I_j^t}^2\rangle-{\langle I_j^t\rangle}^2]}}
\end{equation}
where $\delta$ is the estimation interval, and  $\langle \dots\rangle$ is the sample mean over the detrended AQI series vector $\{I_i^t\}$ of city $i$. In the process of calculating the correlations between two cities' detrended AQI series, we shift the time series backwards and forwards from 1 day to 30 days. Then we calculate the correlation coefficients for the overlap time range of the detrended AQI time series of the two cities, and the largest value is assigned to the Pearson correlation coefficient between the two cities. After we calculate the $N\times N$ correlation coefficients  $C_{ij}$ of each city $i$ and city $j$, we get the air quality correlation matrix.

The $N\times N$ correlation matrix is transformed to network by using the PMFG method \cite{tumminello2005tool}, which is useful and effective to convert the time series data to the complex network, by showing the correlations between different time series datasets. The Pearson correlation coefficients $C_{ij}$ are ranked from the largest one to the smallest one. Then $3(N-2)$ edges are added between cities according to the correlation coefficients rank list (from the largest one), and the new edge adding process should keep the network as a planar graph.

The main advantage of this method is that, at the very beginning, edges are added between cities with larger Pearson correlation coefficients, after many edges are added, the overall picture of the network is almost clear. But there are still a few cities that are not connected to the network, and if new edges are added without the condition of keeping the network as a planar graph, there should be huge number of new edges to be added so as to connect all cities. So with this method, all nodes could be connected to the network with less edges, which makes the correlations between cities clean and clear. Moreover, the minimum spanning tree is more convenient and efficient to show the correlations between different cities. For example, the community structures of networks could be detected with better accuracy.

We have checked the Pearson correlation coefficients distribution of city-pairs with links in the network, normally their Pearson correlation coefficients are larger. This means that the network links in AQNC could reflect the overall picture of the correlations between cities.

\begin{figure}
\subfigcapskip=5pt
\begin{tabular}{cc}
\subfigure[Time evolution of  $\langle AQI\rangle$.]{\label{fig:subfig:a}
\includegraphics[width=0.45\textwidth]{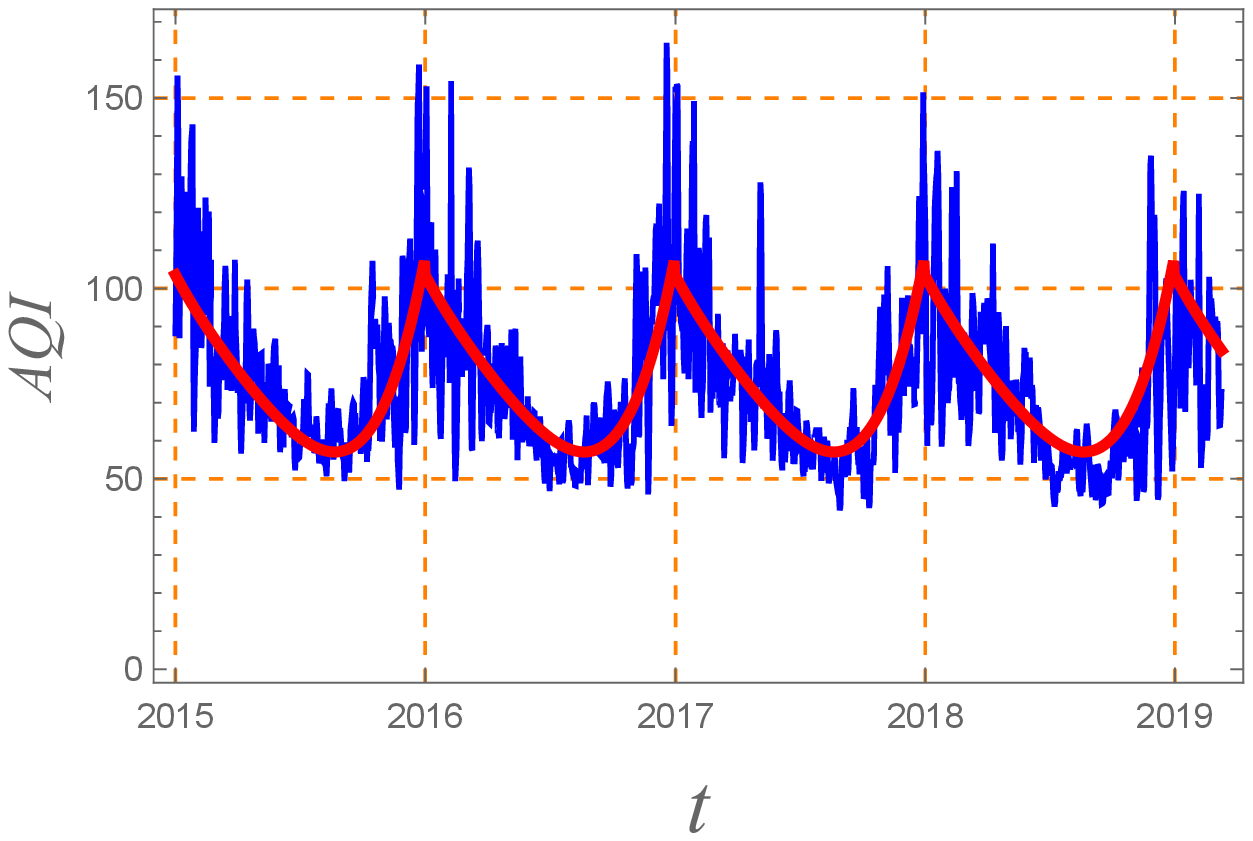}}
&
\subfigure[Time evolution of  $\langle C \rangle$.]{\label{fig:subfig:b}
\includegraphics[width=0.45\textwidth]{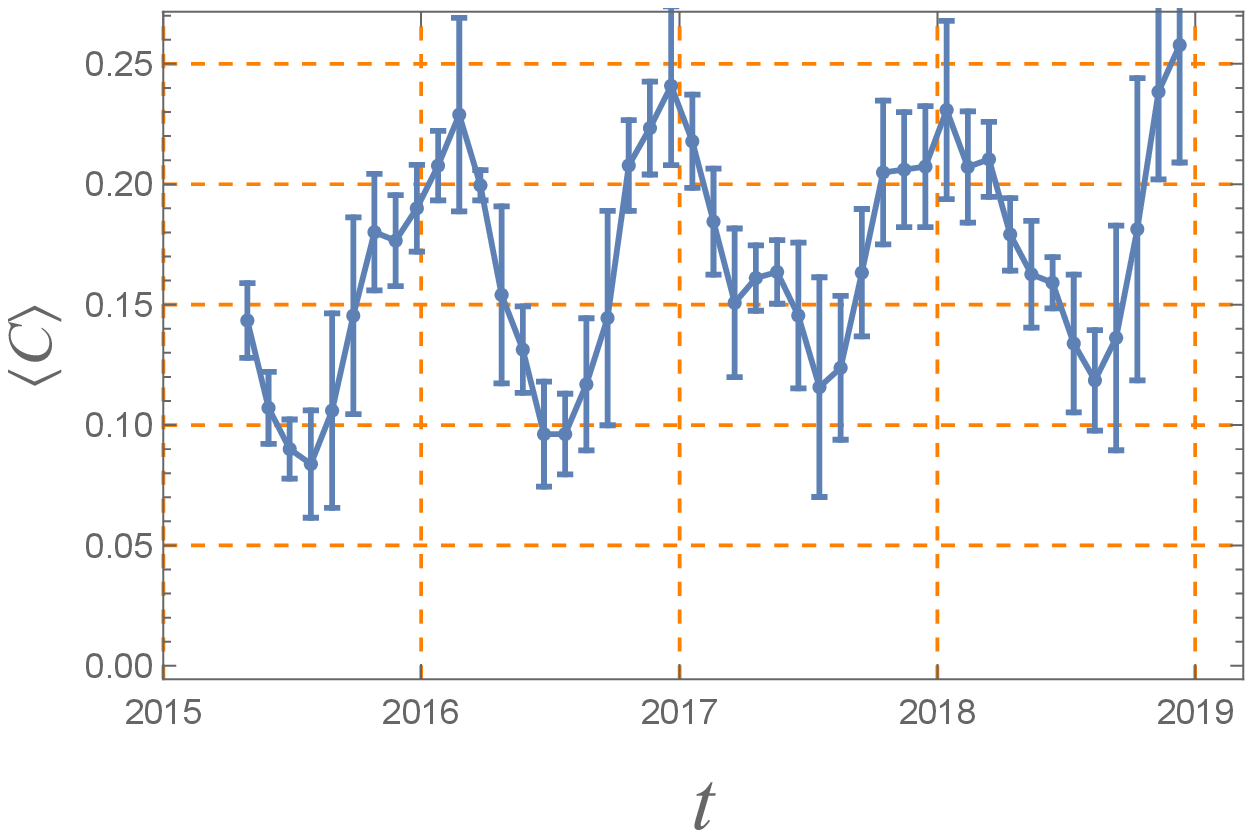}}

\end{tabular}
\caption{(Color online) (a) The time evolution of $\langle AQI\rangle$ (average AQI of 363 cities of China (blue line), the red line just displays the general trend of the evolution of $\langle AQI\rangle$ to guide the eyes.).  (b) The time evolution of  $\langle C\rangle$ (average correlation coefficients $C$ between all cities with different correlation length $\delta$). The error bars represent the standard deviation of $\langle C\rangle$ for different correlation lengths $\delta$ ranging from 30 days to 180 days.}
\label{fig:AQIpearson}
\end{figure}

The average correlation coefficients  $\langle C \rangle$ of each pair of cities are calculated for different correlation length $\delta$ ranging from $30$ days to $180$ days (in order to reduce the fluctuations of only one constant correlation length). The evolutions of $\langle C \rangle$ and the average AQI of all cities are shown in Fig.~\ref{fig:AQIpearson}. When $\langle AQI \rangle$ is larger, $\langle C \rangle$ also becomes larger. While the peaks of the curve of $\langle C \rangle$ slightly shift to the left compared to that of the curve of $\langle AQI \rangle$. It means that after $\langle C \rangle$ reaches the largest value, the air quality would become worse then.  

\subsection{Air pollution diffusion distance}

\begin{figure}
\centering
\begin{tabular}{c}
\includegraphics[width=0.7\textwidth]{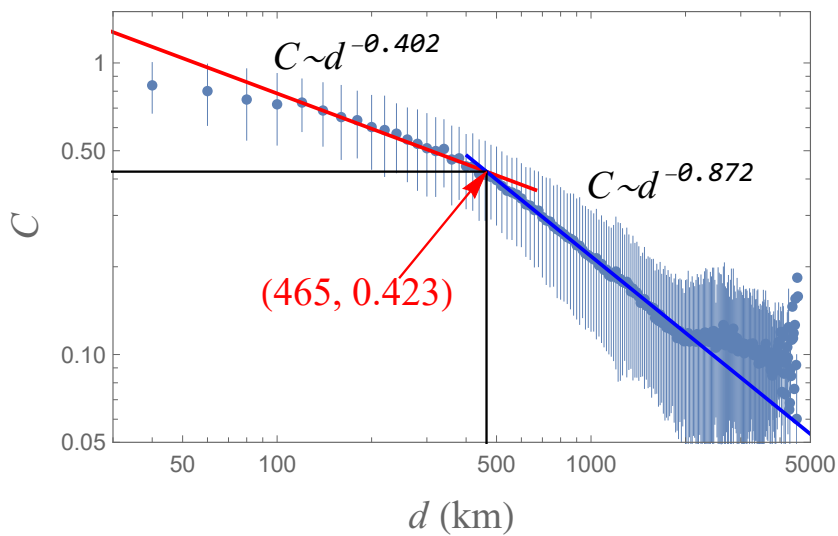}
\end{tabular}
\caption{(Color online) The relationship between the geographical distances $d$ for each pair of cities and their corresponding Pearson correlation coefficients $C$ in double-log scale. The red and blue curves are the power law fittings of the two regimes, the turning pint is $(465, 0.423)$. The correlation coefficients are averaged over each bin of 10 km, the error bars represent the standard deviations of $C$ over each bin.}
\label{fig:pearsondistance}
\end{figure}

Wind is an important factor that affects the diffusion of air pollutions \cite{Zhang2018Correlation}. Normally the closer the geographical distance of two cities, the stronger the correlation of their AQI would be. Therefore, we investigate the relationship between the Pearson correlation coefficients of any two cities and their geographical distances,and plot them in the double-log scale (see Fig.~\ref{fig:pearsondistance}). One could observe that there are two regimes of the relationship, i.e., double power laws with two different slopes. The turning point is ($d=$ 465 km, $C$=0.423). It means that, when $d>465$ km, the correlations between cities would become weaker.
 
\begin{figure}
\centering
\subfigcapskip=5pt
\begin{tabular}{cc}
\subfigure[Geographical distance distribution of any two cities among 363 cities.]{\label{fig:subfigPDF:a}
\includegraphics[width=0.45\textwidth]{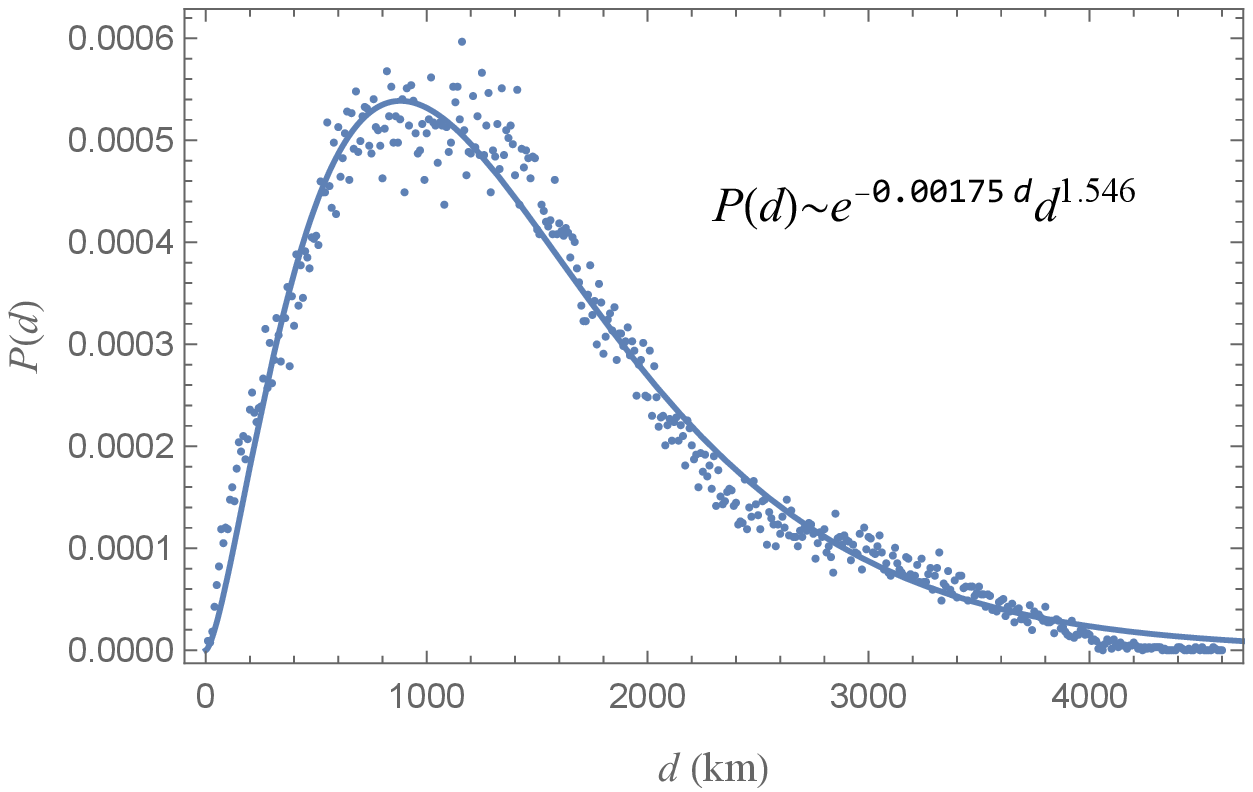}}
&
\subfigure[Geographical distance distribution of the connected cities in AQNC.]{\label{fig:subfigPDF:b}
\includegraphics[width=0.45\textwidth]{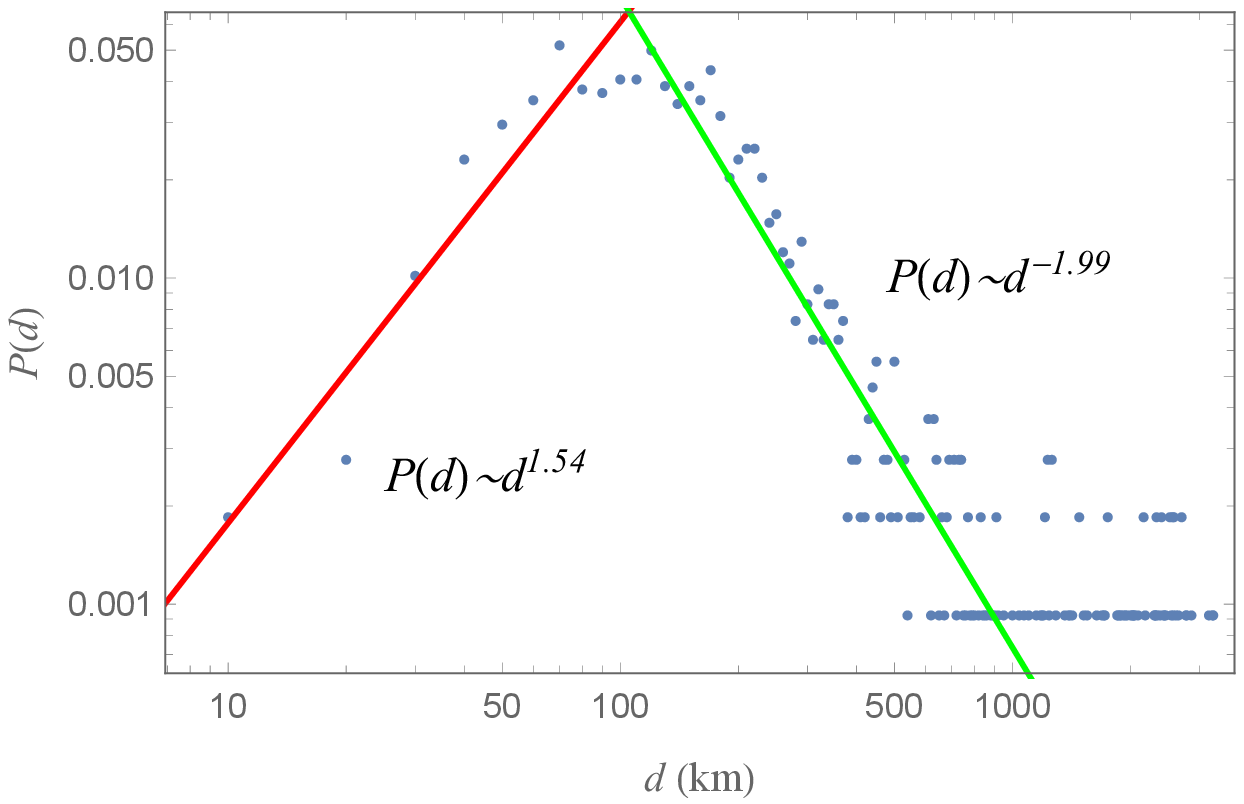}}
\end{tabular}
\caption{(Color online)  (a) The geographical distance distribution of any two cities among 363 cities. The peak of the distribution is around 1000 km. (b) The geographical distance distribution of two cities that have direct links in the network. It is plotted in log-log scale, red line is the fitting cure in the range of $(10, 100)$ km, while the green line is the fitting curve in the range of $(100, 1000)$ km.}
\label{fig:distancedistribution}
\end{figure}

Boers \cite{boers2019complex} studied the global rainfall teleconnections by calculating the probability distribution of distances of links between different places within the network. They found a critical distance of $2500$ km, within the critical distance, the rainfall shows a regional weather system, when the distance is larger than $2500$ km, the rainfall teleconnections are the global-scale ones. In this realm, we both investigate the probability distribution of geographical distances of any two cities and of the connected cities in the AQNC, the results are shown in Fig.~\ref{fig:subfigPDF:a} and (b), respectively. The probability distribution of distances of any two cities follows $P(d)\sim e^{-0.00175d}d^{1.546}$, the peak of the distribution is around 1000 km. But the peak of the probability distribution of distances of links in AQNC is around 100 km. There are two regimes of the distribution, the critical distance is around 100 km. In the region of $(10, 100)$ km, the  probability distribution of distance follows $P(d)\sim d^{1.54}$, while in the region of $(100, 1000)$ km, $P(d)\sim d^{-1.99}$. It means that most of the geographical distances of the connected cities in AQNC are around 100 km. In China, 100 km is more or less the geographical distance between two nearby cities (cities that are in our AQI data list), the air qualities have strong correlations between them. 
Also, comparing Fig.~\ref{fig:subfigPDF:a} (geographical distance distribution of any two cities) and Fig.~\ref{fig:subfigPDF:b}, in the range of [0, 100] km, we observe that both of the two geographical distance distributions have the similar form with $P(d)\sim d^{1.54}$, this indicates that if the geographical distance of two cities is smaller than 100 km, there would be a link between them in the network, so there are strong correlations between them.

\begin{figure}
\centering
\subfigcapskip=5pt
\begin{tabular}{cc}
\subfigure[Average AQI of a single city versus the average AQI of its neighboring cities within 465 km.]{\label{fig:subfigaqisaqia:a}
\includegraphics[width=0.44\textwidth]{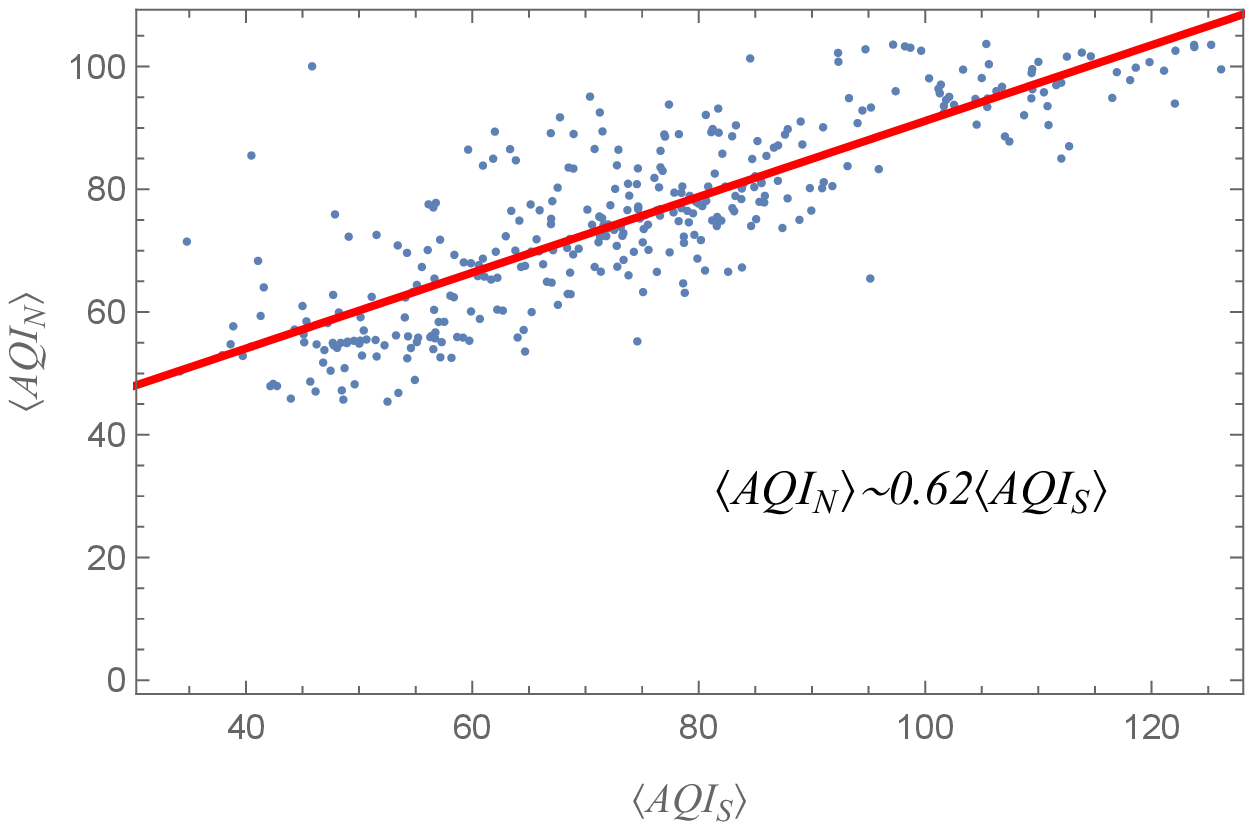}}
&
\subfigure[Probability distribution of the correlation coefficients between each city's $\rm AQI_d$ and its neighboring cities' average $\rm AQI_d$.]{\label{fig:subfigPDFpcc:b}
\includegraphics[width=0.45\textwidth]{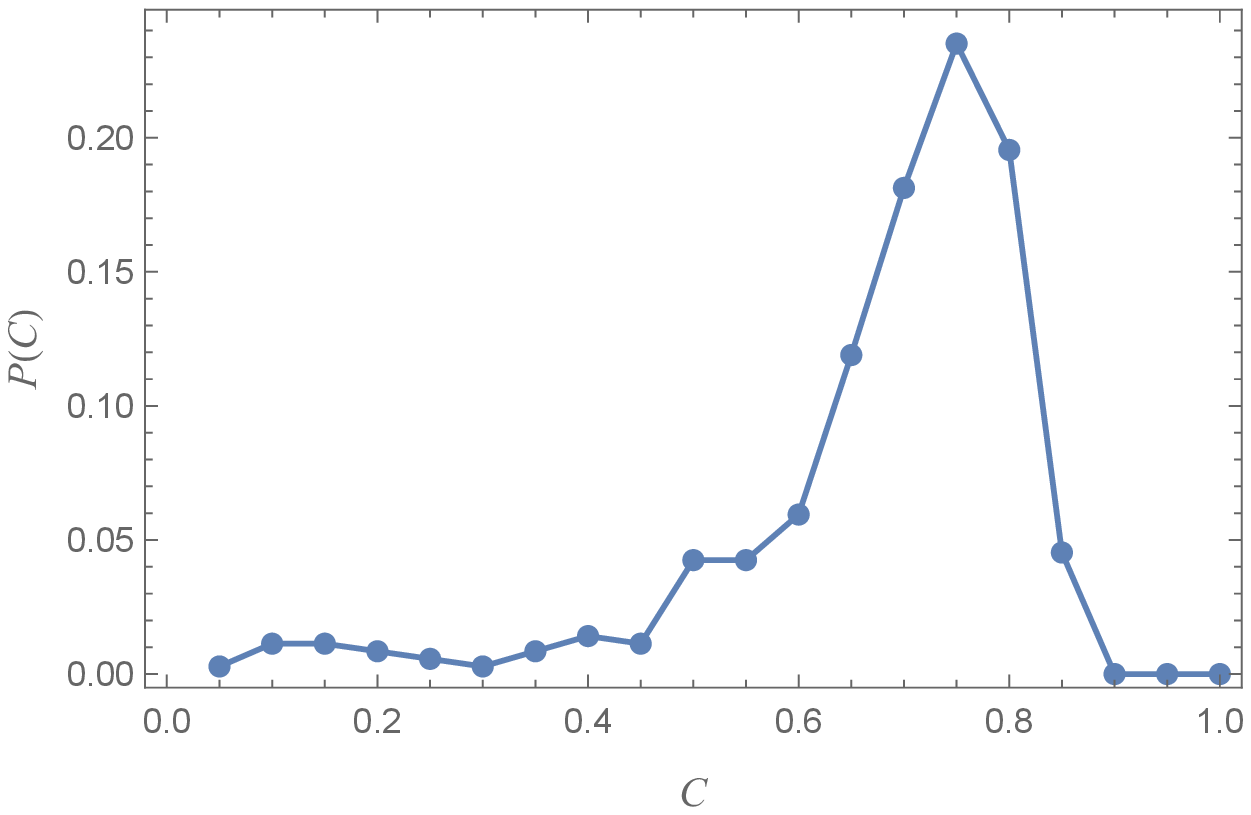}}
\end{tabular}
\caption{(Color online)  (a) The relationship between the average AQI of a single city  $\langle AQI_S \rangle$ and the average AQI of its neighboring cities $\langle AQI_N \rangle$  within $465$ km. (b) The probability distribution of the correlation coefficients between each city's detrended AQI series data and its neighboring cities' average detrended AQI series data.}
\label{fig:aqiareapearson}
\end{figure}

Furthermore, the relationship between the average AQI (over all the periods) of a single city $\langle AQI_S \rangle$ and its neighboring cities' average AQI $\langle AQI_N \rangle$ have been investigated. Considering the relationship between Pearson correlation coefficients and geographical distances of two cities in Fig.~\ref{fig:pearsondistance}, we assign $465$ km as the neighboring influential range. Results are shown in Fig.~\ref{fig:subfigaqisaqia:a}, one can observe that the larger the average AQI of a city, the larger the average AQI of its  neighboring cities. The probability distribution of the correlation coefficients between each city's $\rm AQI_d$ and its neighboring cities' average $\rm AQI_d$ is shown in Fig.~\ref{fig:subfigPDFpcc:b}, it is obvious that the peak of the distribution is around $0.75$, which shows very strong positive correlations. All these results demonstrate that the neighboring cities can have very similar AQI patterns.

\section{Communities of AQNC and their patterns}

In complex network, a community is a group of nodes which are relatively densely connected to each other within the group but sparsely connected to nodes in other groups of the network \cite{porter2009communities}. Detecting communities  \cite{fan2012secom,ratti2010redrawing,tang2012sigma} can not only uncover the correlations between internal structures and functional behaviors of networks, but also have many practical applications in domains such as biology, sociology, economics and climate science, etc.

In our studies, the community means that, the AQI time series of cities within one community have the same pattern, i.e., the evolution features of AQI are similar to each other, so they have strong correlations. The community structure of AQNC is detected based on the edge centrality algorithm \cite{sun2013methods}, the result is compared with the geographical location of cities \cite{zhang2018statistical}. The Hurst exponents of a city's AQI series data and a community's $\langle AQI \rangle$ series data are calculated, respectively. By taking into account the monsoonal distribution, precipitation distribution and other geographical climate factors, we analyze the different patterns of AQI time series belonging to different communities.

\subsection{Results of community detection}

\begin{figure}
\centering
\subfigcapskip=5pt
\begin{tabular}{cc}
\subfigure[Community struture of AQNC.]{\label{fig:community:a}
\includegraphics[width=0.5\textwidth]{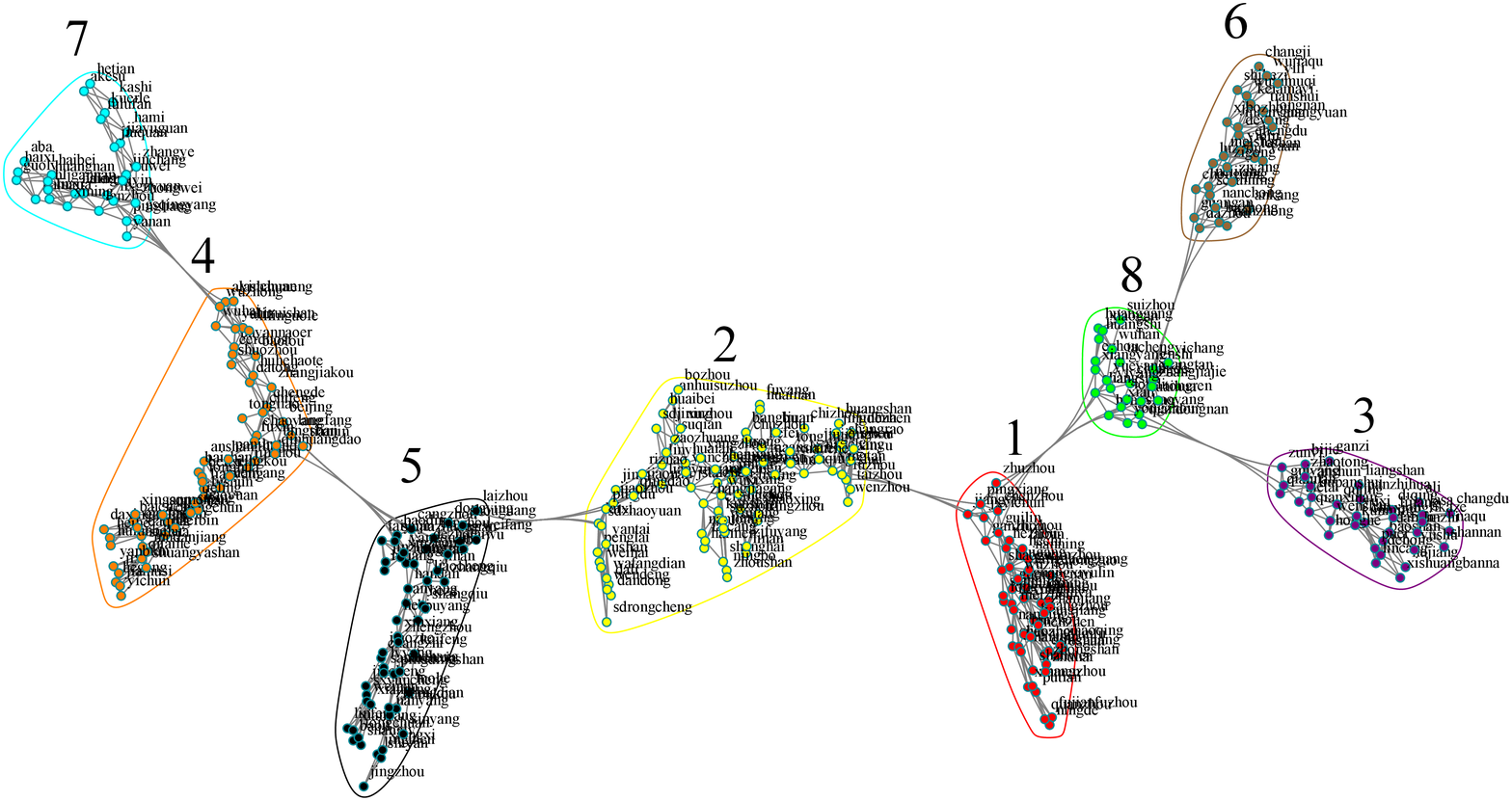}}
&
\subfigure[Geographical locations of cities of different communities.]{\label{fig:community:b}
\includegraphics[width=0.44\textwidth]{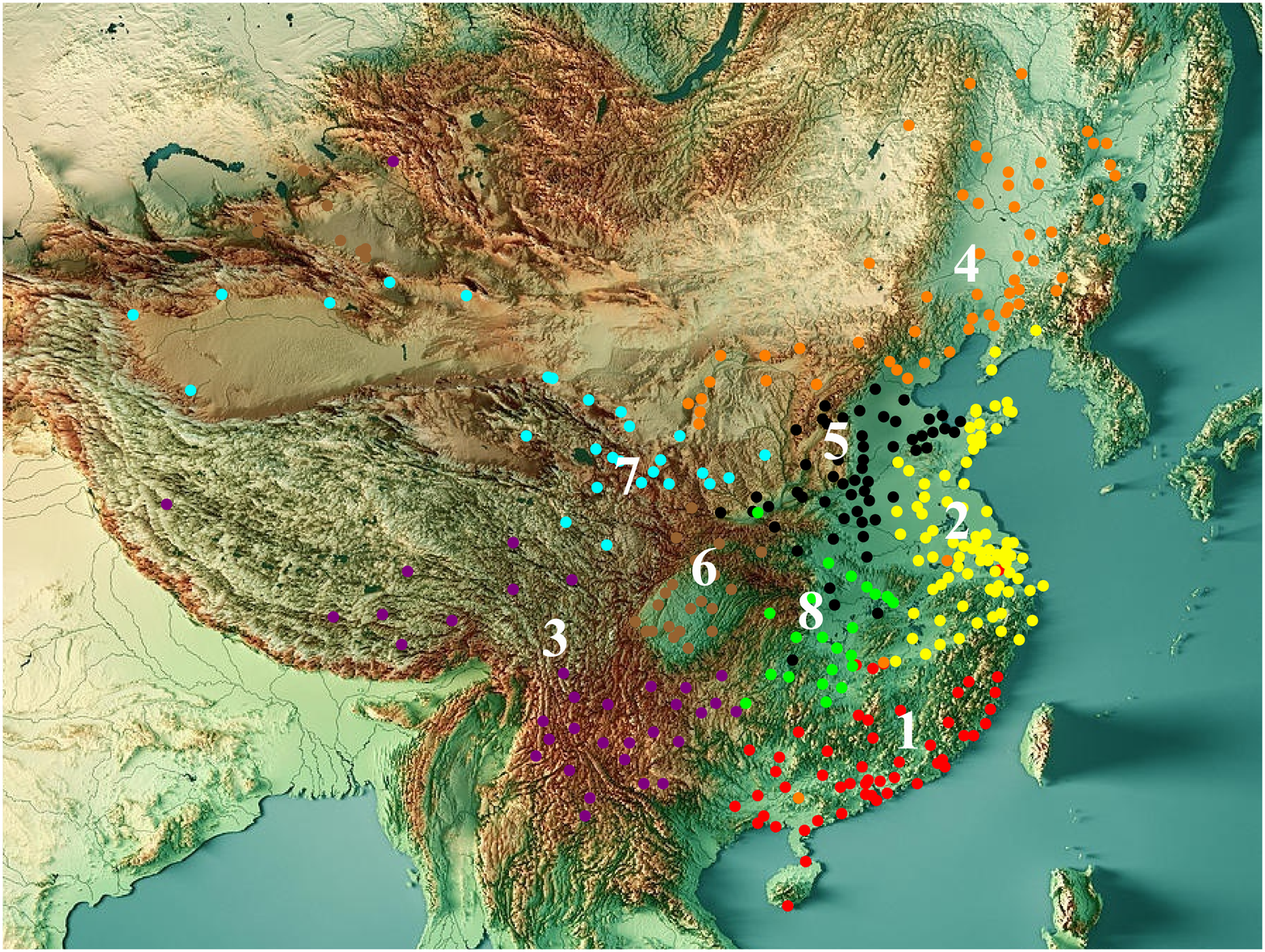}}
\end{tabular}
\caption{(Color online) (a) The community detection result of AQNC. It contains 8 communities and the cities of the same community are denoted by the same color. (b) The geographical locations of cities of different communities on the Chinese map, the colors of cities are the same as shown in (a). The background of figure is the landforms map of China \cite{landformsmap}.}
\label{fig:community}
\end{figure}

As shown in Fig.~\ref{fig:community:a}, eight communities of AQNC have been detected, the value of modularity is 0.807, it means that the community detection result is accurate and reliable, i.e., the AQNC has obvious community structure. From Fig.~\ref{fig:community:b}, one could find that the cities belonging to the same community almost locate in the same region, this indicates that the air quality of a city is largely influenced by its geographical location, i.e., the similar pattern of air quality is mostly due to the similar geographical environment. The locations of the eight communities and their geographical features are as follows \cite{yang2017overview,Sun2016Distribution,landformsmap}:\\
1) {\bf Southeast China}: Area with extreme high precipitation, the cities locate on the monsoonal path.\\
2) {\bf East China}: Plain area with medium precipitation, close to the East China Sea. \\
3) {\bf Southwest highland of China}: Highland area with rough terrain, medium precipitation and low wind. \\
4) {\bf Northeast China}: Plain area with low precipitation and high wind. \\
5) {\bf Central-north China}: Plain area with medium precipitation and low wind.\\
6) {\bf Basin of China}: Basin area with medium precipitation and low wind. \\
7) {\bf Gobi desert of China}: Gobi desert area with rough terrain, low precipitation and high wind.\\
8) {\bf Central-south China}:  Plain area with high precipitation and medium wind.

\subsection{AQI pattern analysis by Hurst exponent and environmental factors}
\begin{figure}[htbp]
\centering
\subfigure[Hurst exponents of 363 cities.]{
    \label{fig:hurstdistribution}
   \includegraphics[width=0.45\textwidth]{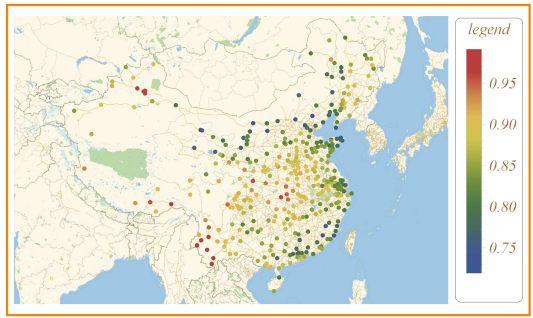}}
     \hspace{0.25in}
  \subfigure[Hurst exponents of 8 communities.]{
    \label{fig:averagehurstdistribution}
   \includegraphics[width=0.45\textwidth]{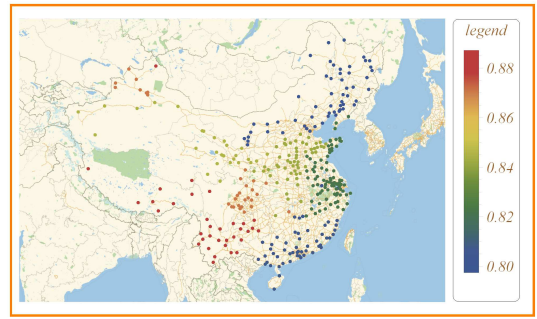}}\\
  \subfigure[The average AQI of 363 cities.]{
    \label{fig:averageaqidistribution}
   \includegraphics[width=0.45\textwidth]{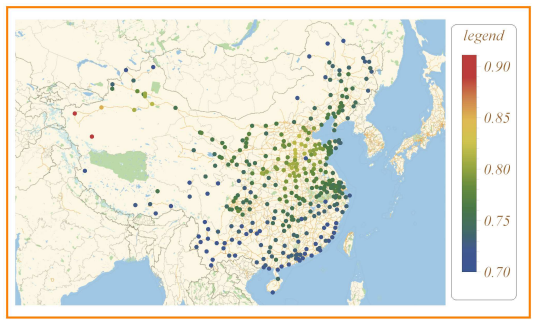}}
     \hspace{0.25in}
  \subfigure[Community versus geographical environment.]{
    \label{fig:Chinaprecipitationandmonsoonalcommunity}
   \includegraphics[width=0.45\textwidth]{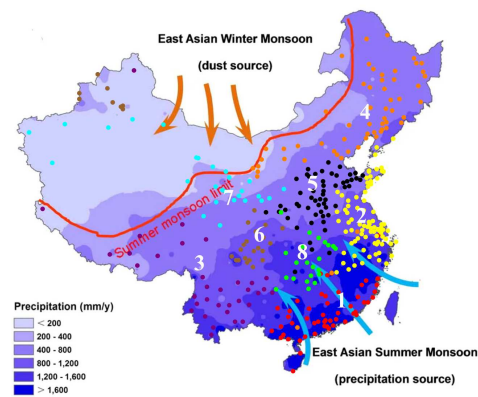}}
  \centering
  \caption{(Color online) The heat map of the Hurst exponents of (a) 363 cities and (b) 8 communities. (c) The heat map of the average AQI of the 363 cities. (d) The geographical distribution of precipitation and monsoon versus communities (the background map is cited from \cite{Sun2016Distribution}).}
  \label{fig:Hurstandwind}
\end{figure}

The Hurst exponent is employed as a measure of the long-term memory of time series. It relates to the autocorrelations of the time series, and the rate at which these decrease as the lag between pairs of values increases \cite{Hurst1951Long}. We calculate the Hurst exponent by using the rescaled range analysis (R/S analysis) \cite{bassingthwaighte1994evaluating}, for a time series of length $n$, $X=X_1, X_2,..., X_n$, the rescaled range is calculated as follows:
\begin{equation}
m=\frac{1}{n}\sum_{i=1}^nX_i,
\end{equation}
\begin{equation}
Y_t=X_t-m \qquad \textrm{for} \quad t=1,2,...,n,
\end{equation}
\begin{equation}
Z_t=\sum_{i=1}^tY_i \qquad \textrm{for} \quad t=1,2,...,n,
\end{equation}
\begin{equation}
R(n)=\max(Z_t)-\min(Z_t)\qquad \textrm{for} \quad t=1,2,...,n,
\end{equation}
\begin{equation}
S(n)=\sqrt{\frac{1}{n}\sum_{i=1}^n(X_i-m)^2}
\end{equation}
where $m$ is the mean value of the series data, $Y_t$ is the mean-adjusted series, $Z_t$ is the cumulative deviation series, and $R(n)$ represents the range of deviation, $S(n)$ is the standard deviation (normalization factor). Using the above formulas, we can calculate the rescaled range $R/S(n)=R(n)/S(n)$, averaging all the partial time series by length $n$, and obtain $R/S(n)\sim n^{H}$, in which $H$ is the Hurst exponent. In the calculation process for Hurst exponent, we found that the R/S line becomes not smooth when $n>365$, so we set one year as the upper bound length of long-term memory. Values of Hurst exponents of the AQI time series of all 363 cities are shown in Fig.~\ref{fig:hurstdistribution}, and the value of the Hurst exponent $H_A$ of the average AQI time series of all cities is 0.955. These results indicate that the AQI time series have the strong long-term memory effects, regardless of  the climate patterns or geographical locations of the cities. 

\begin{figure}
\centering
\begin{tabular}{c}
\includegraphics[width=0.7\textwidth]{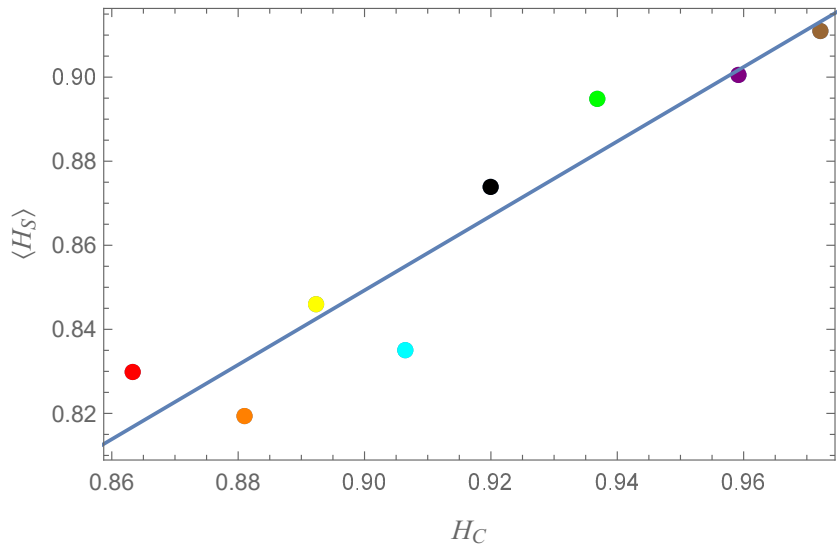}
\end{tabular}
\caption{(Color online)  The relation of a community's Hurst exponent $H_C$ versus the average Hurst exponent of all single cities within the community $\langle H_S \rangle$. As can be seen, they are linearly correlated, except community $4$ with orange color and community $7$ with cyan color.}
\label{fig:hurstregionalcompare}
\end{figure}

We average the AQI time series of cities within the same community, and calculate each community's Hurst exponent denoted as $H_C$ (see Table.~\ref{table:areacompare} and Fig.~\ref{fig:averagehurstdistribution}). The community's Hurst exponent $H_C$ was compared with the average Hurst exponent of all single cities within the community (denoted as $\langle H_S \rangle $) in Fig.~\ref{fig:hurstregionalcompare}. The $H_C$ is larger than $\langle H_S\rangle$ for each community, since the time series of average AQI of a community has smaller fluctuations. $H_C$ and $\langle H_S\rangle$ are linearly correlated except community 4 (Orange color point) and community 7 (Cyan color point) that exhibits an abnormal feature (see Fig.~\ref{fig:hurstregionalcompare}). The Hurst exponents of different communities posses their own features, since for each community, it has its own geographical aggregation effect. The datasets of the precipitation, monsoon, and wind energy data of different areas are obtained from the previous studies and literatures \cite{yang2017overview,Sun2016Distribution,landformsmap}. Thus for example we roughly divide the precipitation into four levels according to their values, and got the levels of low, medium, high and extreme high. The geographical characteristics of each community, including the precipitation, monsoon, wind, regional geomorphic feature and average Pearson correlation coefficient  $\langle C_c\rangle$, are presented in Table.~\ref{table:areacompare}.

\begin{table}
\centering
\caption{\label{table:areacompare}
Basic quantities and features of the $8$ communities. $\langle AQI\rangle$ is the average AQI of a community through the whole period. $H_C$ is the Hurst exponent of a community. $\langle H_S \rangle$ is the average Hurst exponent of all single cities within the community. $\langle C_c\rangle$ is the average Pearson correlation coefficient among cities of the same community with distance being less than 465 km.}
\footnotesize
\begin{tabular}{m{2.0cm}<{\centering}m{1cm}<{\centering}m{1cm}<{\centering}m{1.0cm}<{\centering}m{1.0cm}<{\centering}m{1.0cm}<{\centering}m{2.2cm}<{\centering}m{3.5cm}<{\centering}}
\br
{\bf Community label}&Color&$\langle AQI\rangle$&$H_C$&$\langle H_S \rangle$&$\langle C_c\rangle$&Precipitation&Geographical features\\ \hline
1&Red&54.46&0.863&0.83&0.630&Extreme High&Close to ocean, in monsoon path \\
2&Yellow&74.99& 0.892& 0.846&0.576&Medium&Close to ocean,  not in monsoon path\\
3&Purple& 49.84&0.959&0.901&0.424&Medium&Highland, low wind\\
4&Orange& 73.04&0.881&0.819&0.569&Low&Plain, high wind\\
5&Black&103.37& 0.92&0.874&0.558&Medium&Plain, low wind\\
6&Brown&76.09&0.972&0.911&0.625&Low&Basin, low wind\\ 
7&Cyan&88.46&0.906&0.835&0.481&Low&Gobi desert, high wind\\
8&Green&75.25&0.937&0.895&0.659&High&Plain, medium wind\\
\br
\end{tabular}
\end{table}

From the community analysis of AQNC, we can discuss the results as follows: 1) The values of the $\langle H_S \rangle$ and $H_C$ are relatively large in the basin area, due to the very low wind/air exchanges, which implies that the AQI time series of the basin area has strong long-term memory. 2) The wind and precipitation are negatively correlated with the Hurst exponents of single cities, as well as communities. Since the wind and precipitation could influence very large area, so as to break the air pollution accumulative effect \cite{kirkby1987hurst}, and lessen the Hurst exponents. 3) The highland and gobi desert areas have rough terrain which can lead to low correlation between cities (small $\langle C_c\rangle$), and large $H_C$, such as communities 3 and 7. The community 7 have high wind that leads to small $\langle H_S \rangle$, but rough terrain reduces the air flow and leads to large $H_C$. 4) The wind (local area) could reduce the value of the Hurst exponent of single cities, but reduce the value of communities ($H_C$) slightly. Since it can only influence the AQI change in smaller area (single city), and it is not the same as monsoon which can influence large area (the community area), such as community 4.

\section{Motifs of AQI time series by the visibility graph}

The visibility graph method was used to analyze evolution features, especially the motifs of  AQI time series. Some standard motifs were recognized by comparing the original time series with the shuffled ones. We further studied the stability of these motifs of different communities by checking their long-term memory effects.
 
\subsection{The process and properties of visibility graph}

\begin{figure}[htbp]
\centering
\subfigure[Time series.]{
    \label{fig:VisibilityGraphtransform}
   \includegraphics[width=0.35\textwidth]{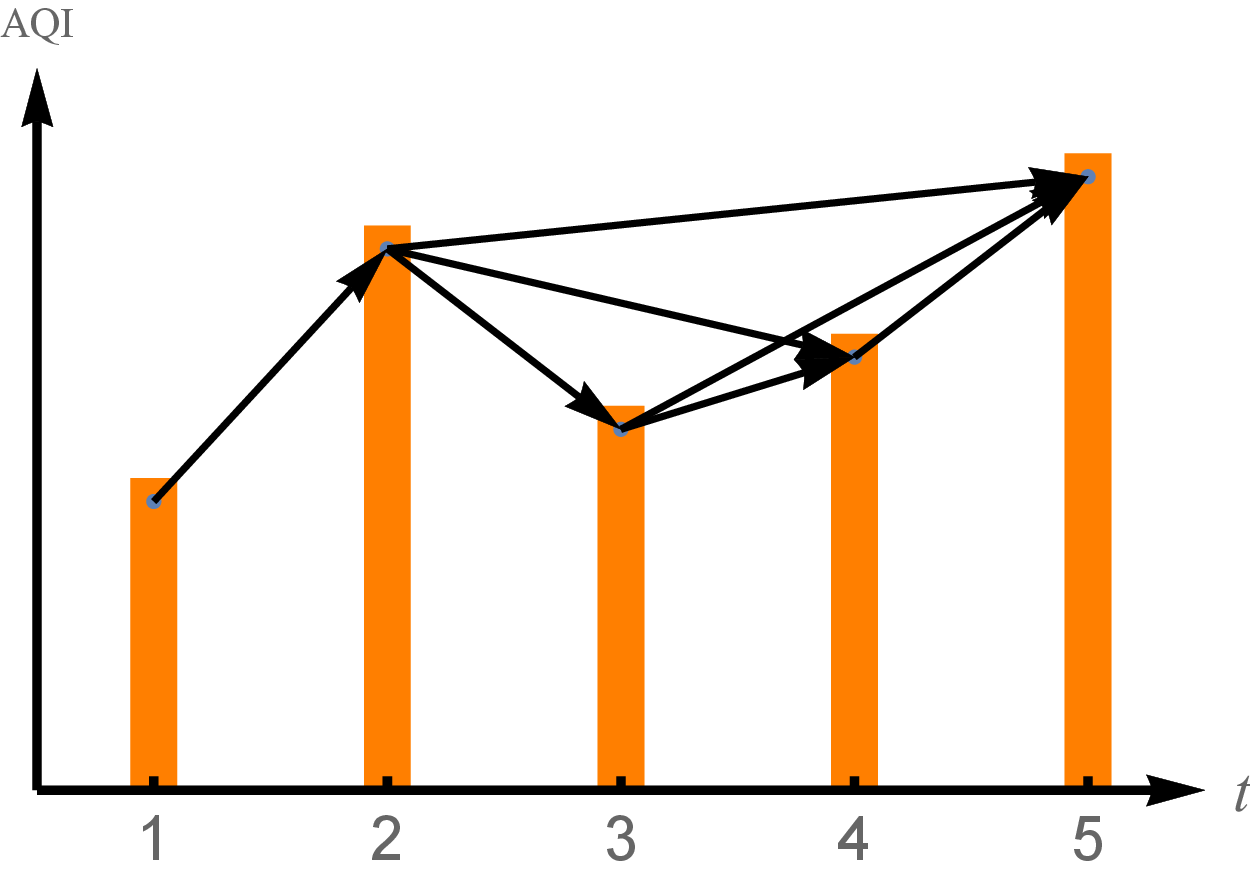}}
  \subfigure[Visibility graph mode.]{
    \label{fig:VisibilityGraphmode}
   \includegraphics[width=0.25\textwidth]{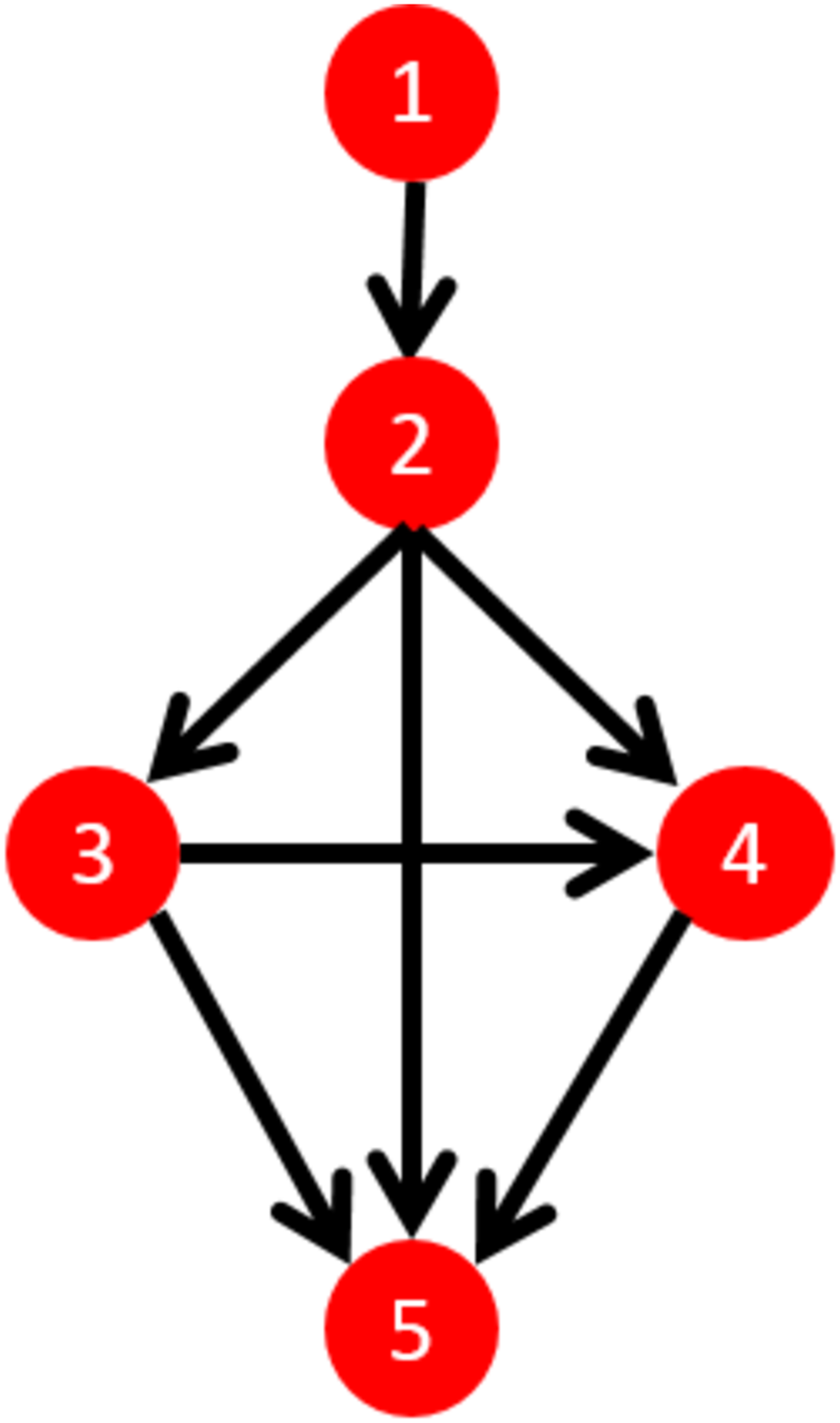}}
  \subfigure[Adjacent matrix.]{
    \label{fig:Visibilitymatrix}
   \includegraphics[width=0.35\textwidth]{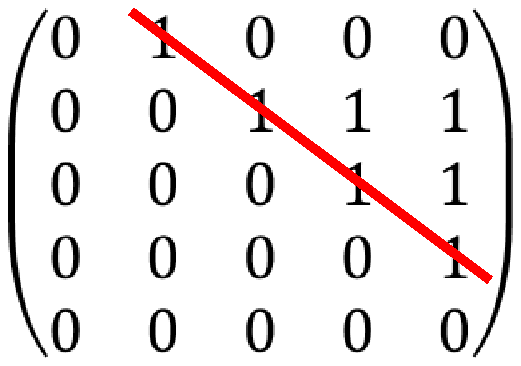}}

  \centering
  \caption{(Color online) The process of transforming the time series to visibility graph and its corresponding adjacent matrix.}
  \label{fig:Visibilityprocess}
\end{figure}

The visibility graph was proposed by Lacasa \cite{lacasa2008time} and used to transform the time series to graphs, then these graphs are regarded as vertices to construct a new network \cite{Mutua2016Visibility}. Here, we choose five timing points from $t$ to $t+4$ and set every timing point as one vertex in the network, if two elements (e.g., with height being the value of AQI) at different timing points are visible, they will be connected in the network (see Fig.~\ref{fig:VisibilityGraphtransform}). So we considered each time window with length 5, and set every timing point as a vertex. An edge is added if two vertices (with the height of AQI value) could see each other, for which the direction is from the early timing vertex to the subsequent one. Please find the schematic diagram in Fig.~\ref{fig:Visibilityprocess}. Fig.~\ref{fig:Visibilitymatrix} is the corresponding adjacency matrix of the graph in Fig.~\ref{fig:VisibilityGraphmode}. The values of elements on the red line are $1$ since they reflect the adjacent timing points. The values of elements below the diagonals are $0$. Thus we can represent the visibility graph by the elements above the red line. According to the previous results \cite{stephen2015visibility}, for 5 timing points, the connection patterns in the visibility graph can be divided into 25 kinds. 

\begin{figure}
\centering

\begin{tabular}{cc}
\subfigcapskip=5pt
\subfigure[Southeast China.]{
\includegraphics[width=0.4\textwidth]{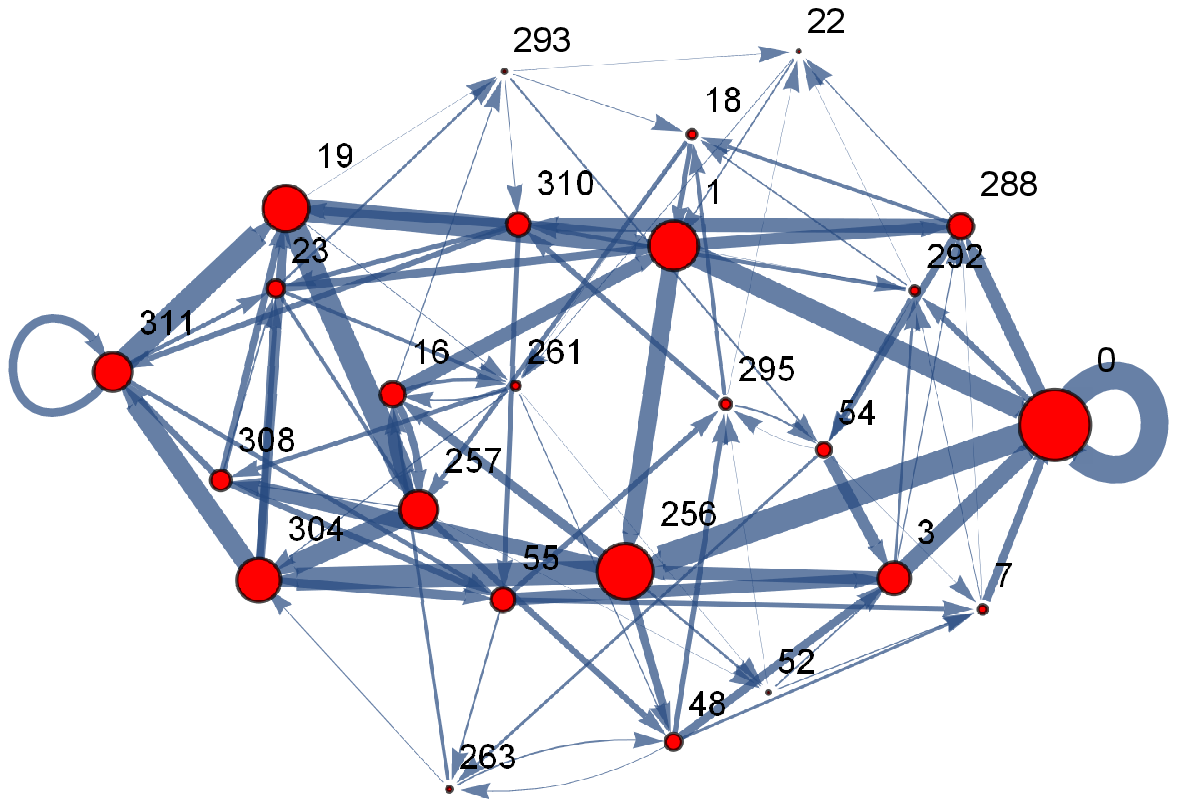}}
&
\subfigure[East China.]{
\includegraphics[width=0.35\textwidth]{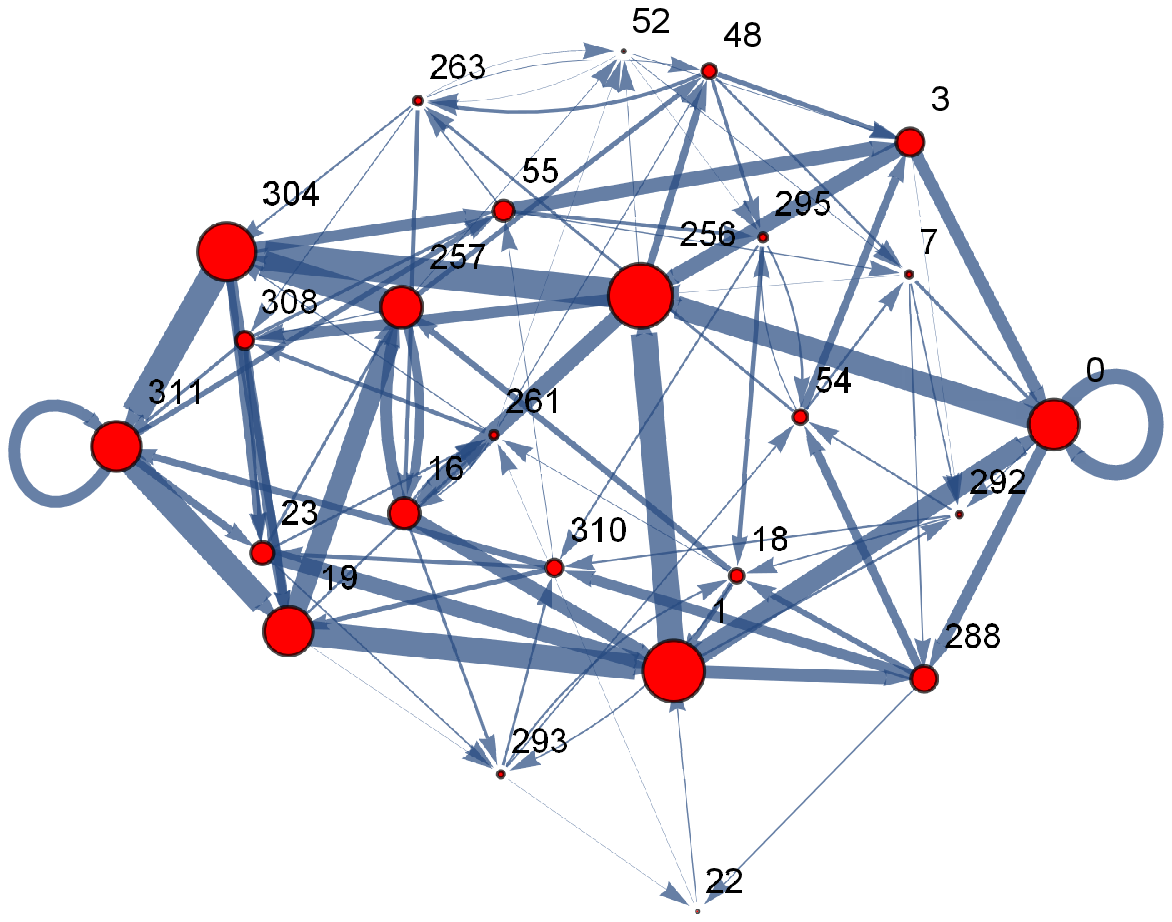}}
\\
\subfigure[Southwest highland of China.]{
\includegraphics[width=0.35\textwidth]{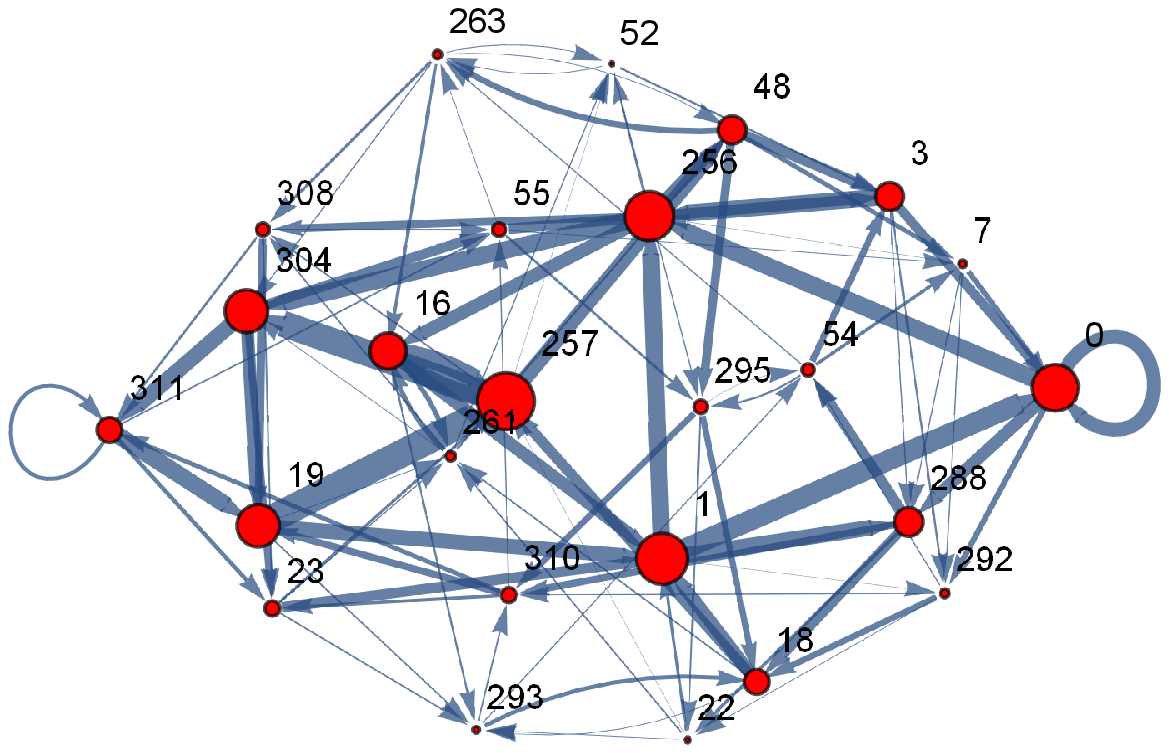}}
&
\subfigure[Northeast China.]{
\includegraphics[width=0.35\textwidth]{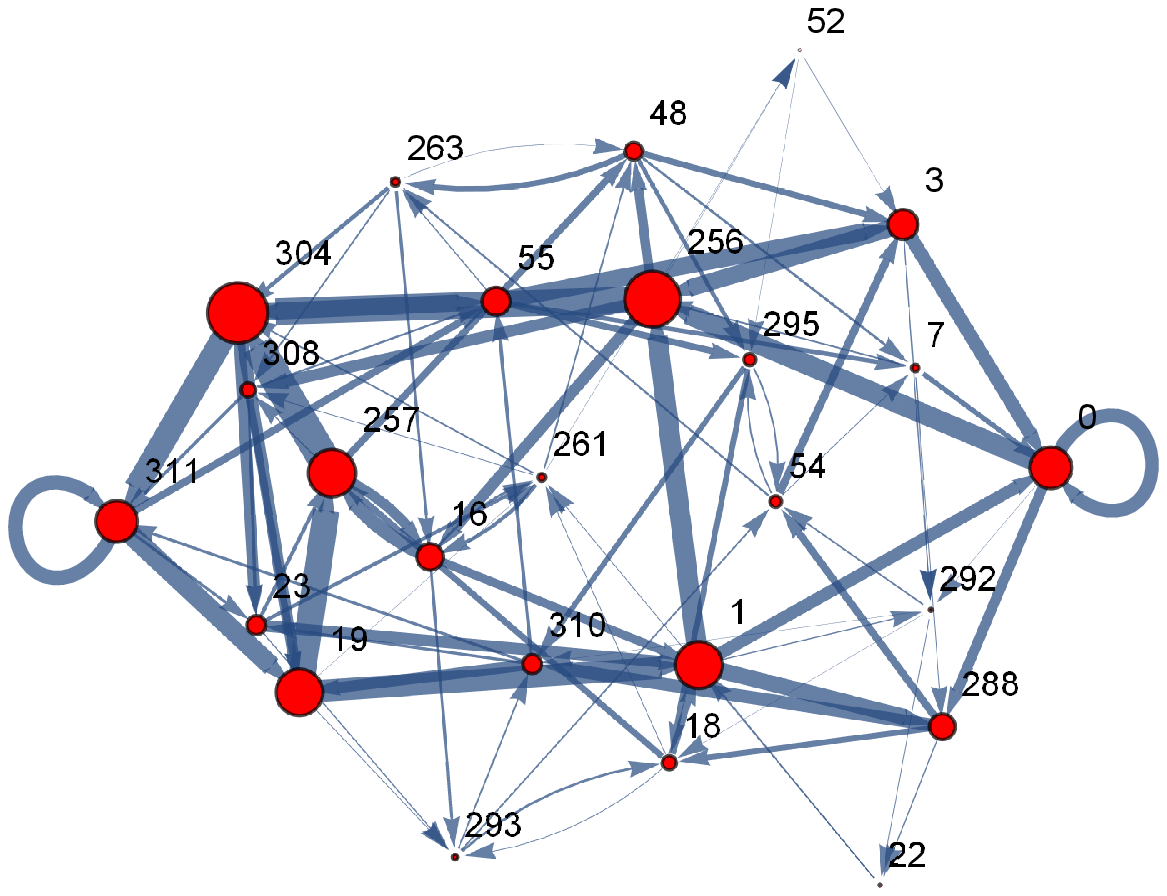}}
\\
\subfigure[Central-north China.]{
\includegraphics[width=0.35\textwidth]{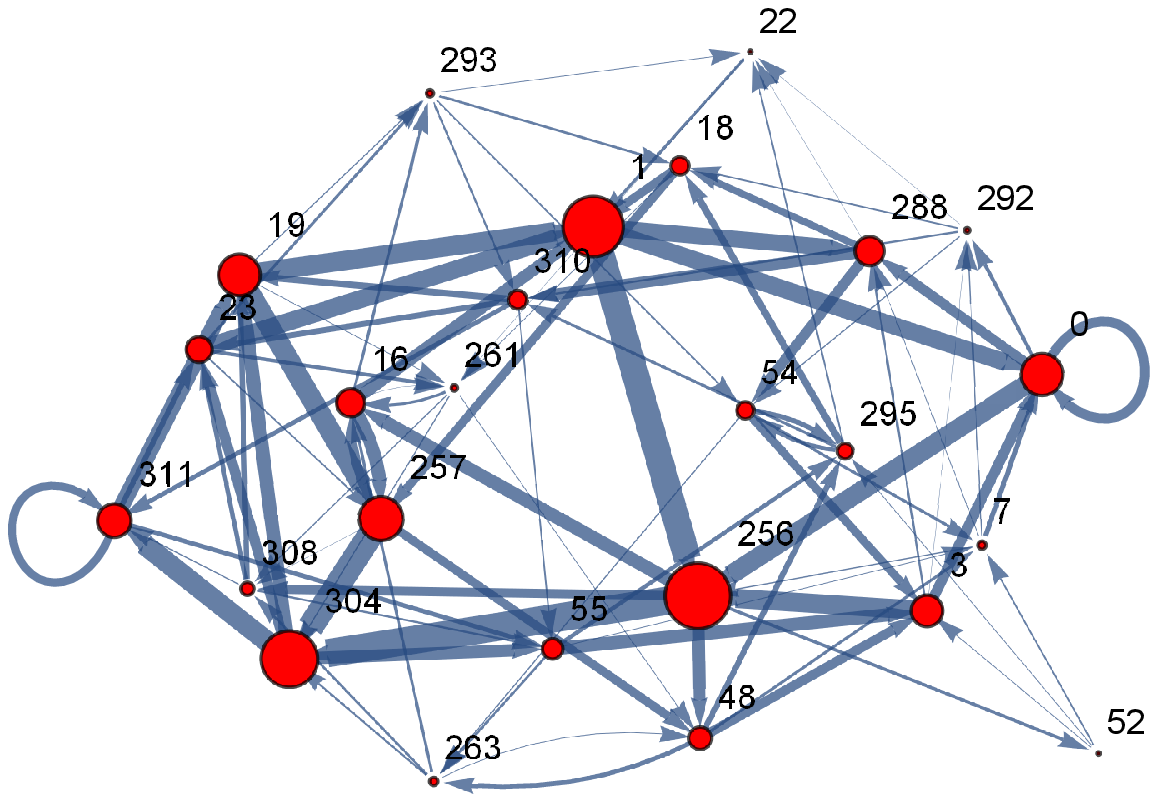}}
&
\subfigure[Basin of China.]{
\includegraphics[width=0.35\textwidth]{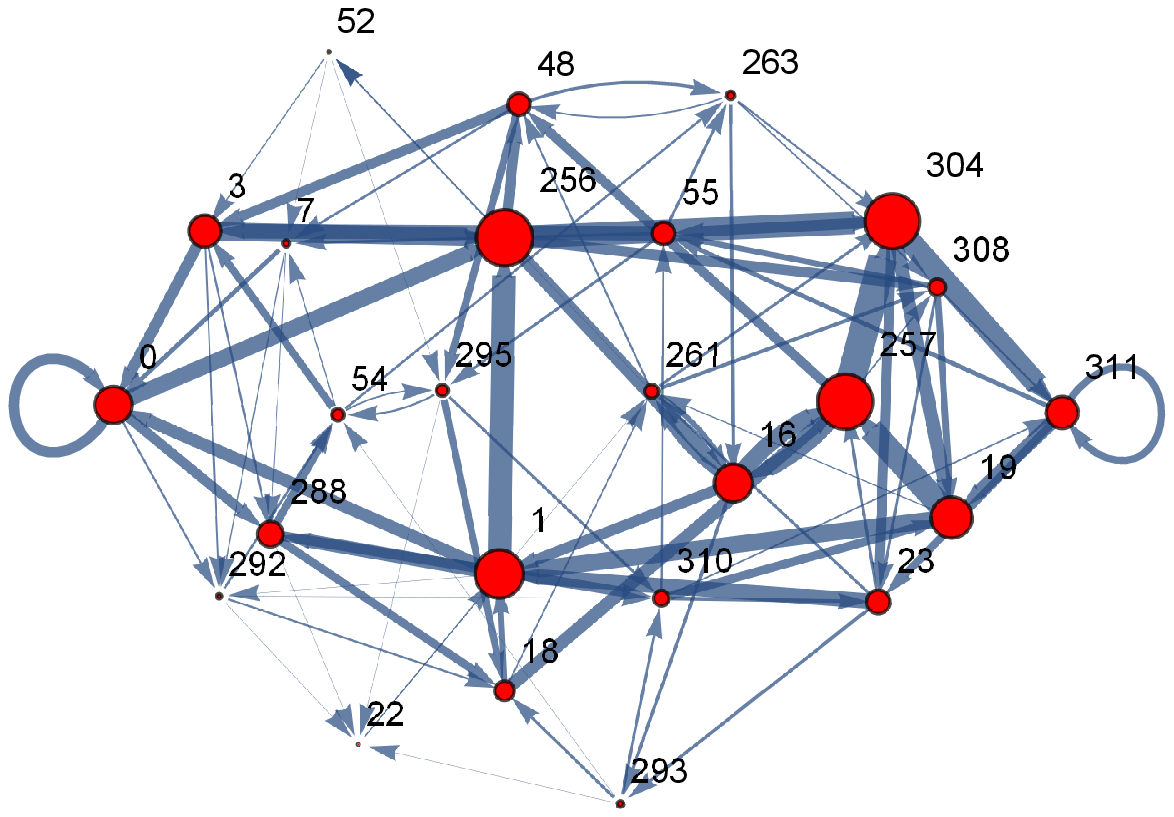}}
\\
\subfigure[Gobi desert of China.]{
\includegraphics[width=0.35\textwidth]{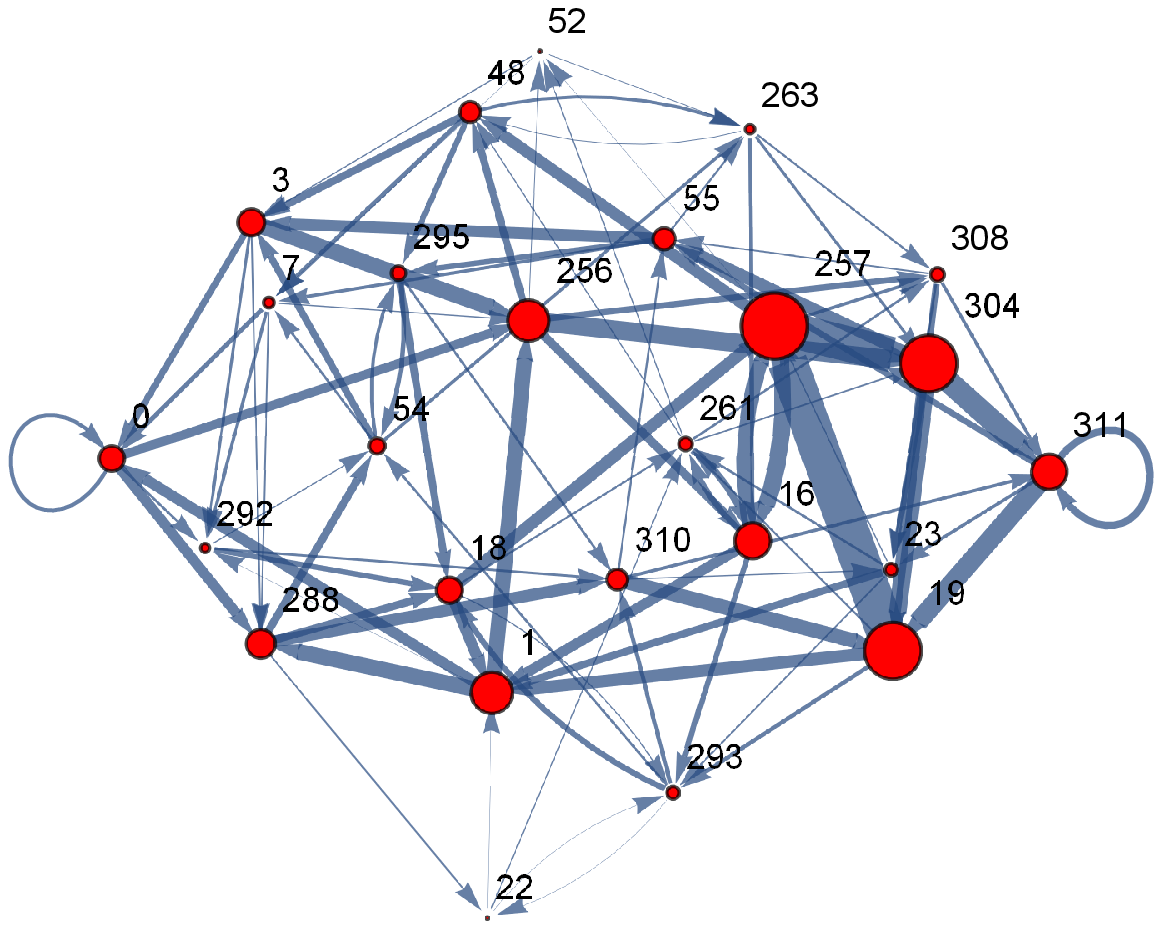}}
&
\subfigure[Central-south China.]{
\includegraphics[width=0.35\textwidth]{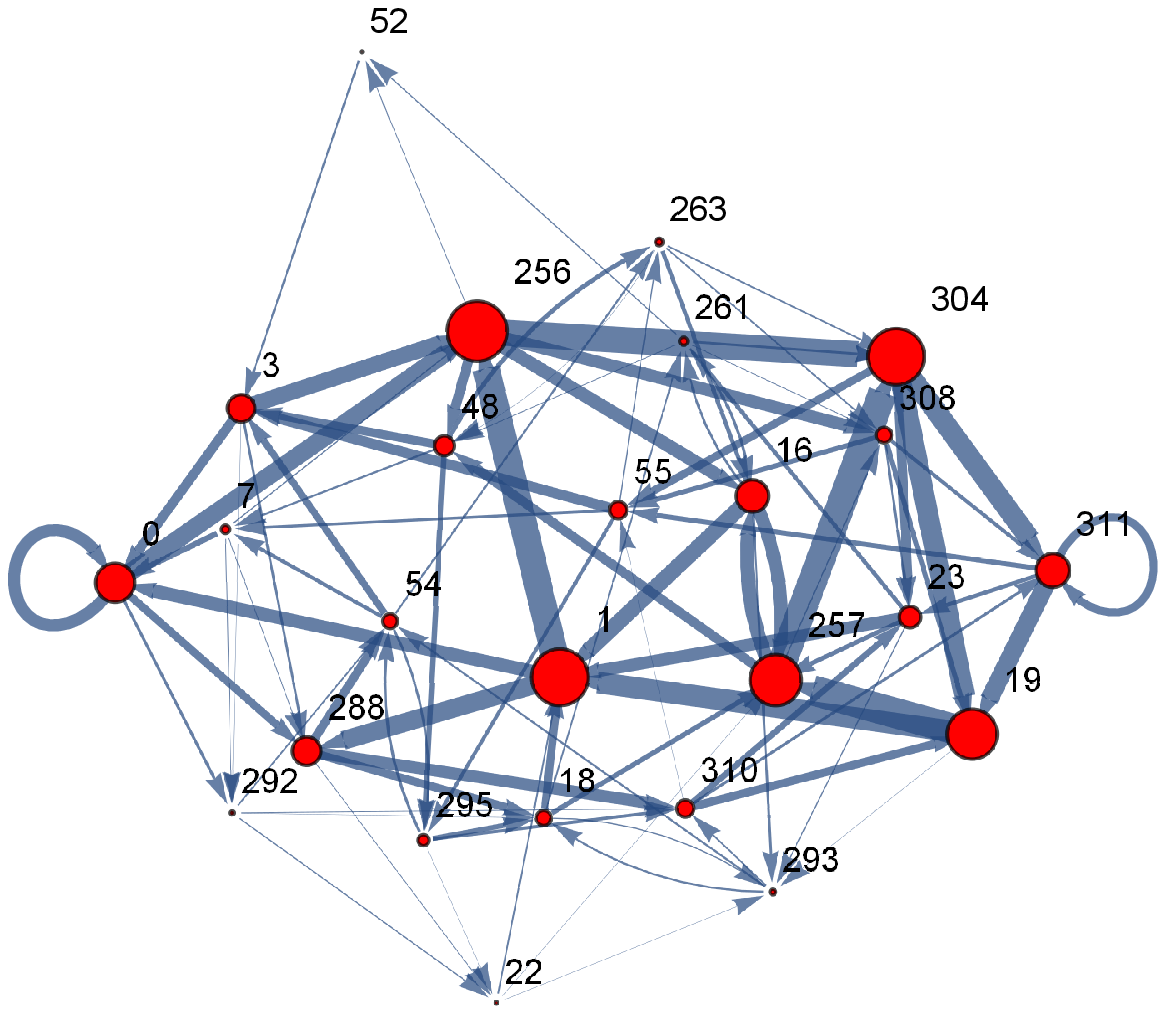}}
\end{tabular}

\caption{(Color online)  The transforming patterns of the visibility graphs of eight communities, the thickness of the line is proportional to the frequency of that transformation, and the size of the vertex is proportional to the frequency of the visibility graph in the AQI time series. The labels of the visibility graphs are named in the way that, if we transform that label to the binary number with nine-digits (0 or 1), then these nine-digits (0 or 1) just correspond to the nine numbers at the top right-hand corner of the adjacent matrix of visibility graph.}

\label{fig:1visibilitygraph}
\end{figure}

We denote the network composed of 5 timing vertices as a mode $g_t$, and the next shifted (only one step) 5 timing vertices as $g_{t+1}$, and so on. So the time series of AQI can be denoted as $g_t, g_{t+1}, g_{t+2}, ... ,g_{t+n}$. For the next step, we add an edge for any two consecutive mode $g_t, g_{t+1}$, and the same mode of visibility graph as one point, then the network is constituted by 25 visibility graphs (most visibility graphs can't be generated), and the mutual transformation can be obtained, which is named as the visibility network. The network is shown in Fig.~\ref{fig:1visibilitygraph}, which could reflect the patterns and rules of AQI evolution. Each node corresponds to a visibility graph. The digit next to the node is the name of the visibility graph, in the way that, if we transform that name-digit to the binary number with nine-digits (0 or 1), then these nine-digits (0 or 1) just correspond to the nine numbers at the top right-hand corner of the adjacent matrix of visibility graph. So the name-digit could reflect the very basic information of the visibility graph. The larger points imply the patterns appear in AQI series much more frequently, and the thicker line means the two patterns have higher probability of transforming from one to another. We calculate the top 8 nodes (visibility graphs) with large frequencies of all communities (see Table.~\ref{table:visibilityg8}).

\begin{table}[hb]
\caption{The top 8 large frequency visibility graphs and corresponding time series of eight communities. The labels of the visibility graphs are the same as in Fig.~\ref{fig:1visibilitygraph}.}
\label{table:visibilityg8}
\footnotesize
\begin{tabular}{ccccc}
\br
Graph label&256&304&1&257\\ 
\hline
Visibility graph&
\begin{minipage}{0.17\textwidth}
\includegraphics[width=0.9\textwidth]{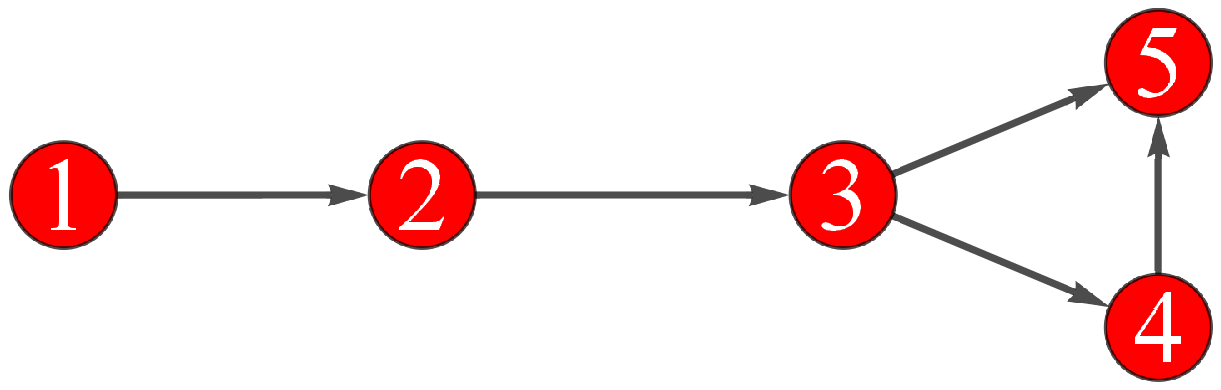}
\end{minipage}&\begin{minipage}{0.15\textwidth}
\includegraphics[width=0.9\textwidth]{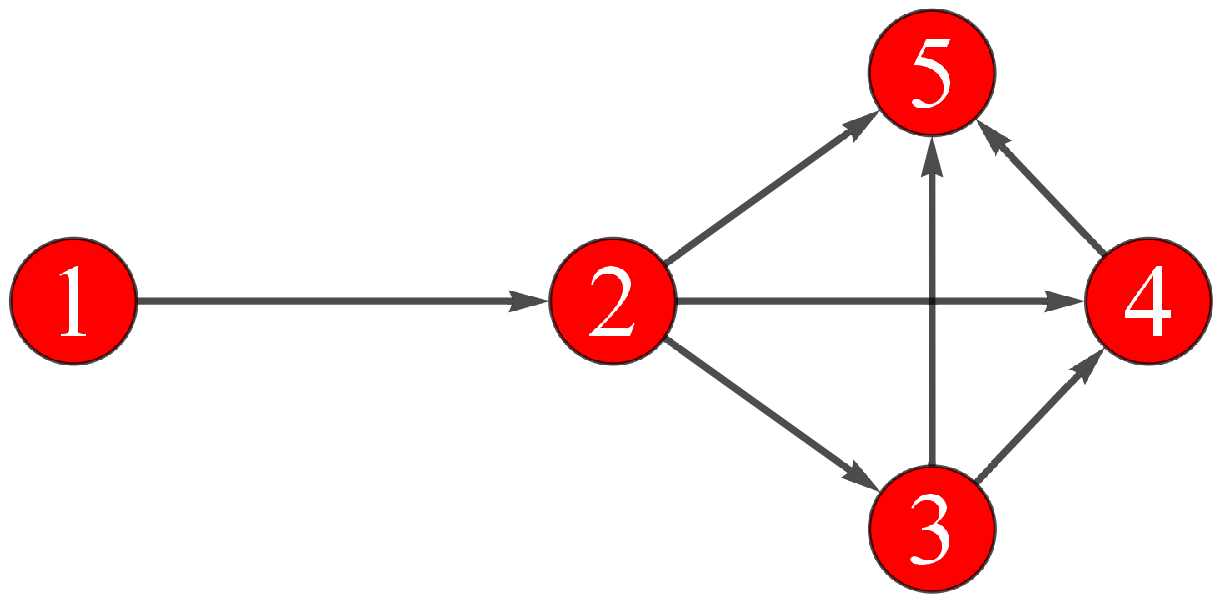}
\end{minipage}&\begin{minipage}{0.15\textwidth}
\includegraphics[width=0.9\textwidth]{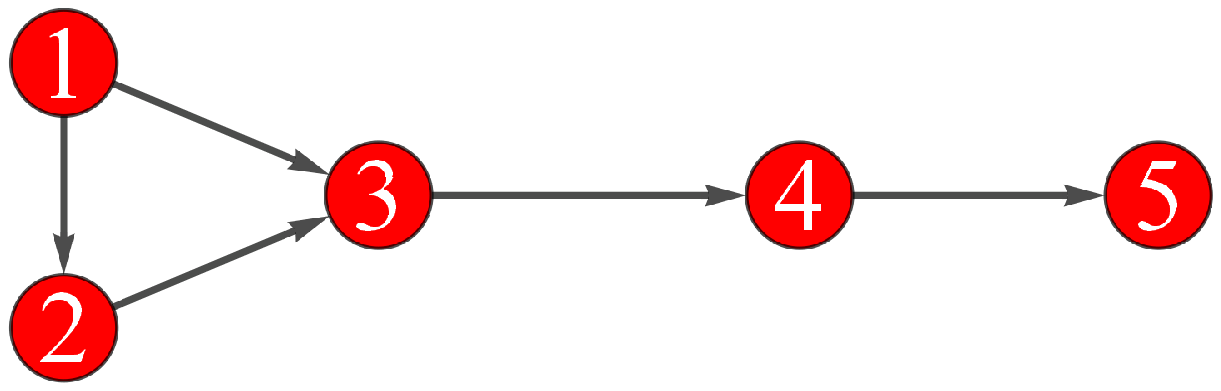}
\end{minipage}&\begin{minipage}{0.15\textwidth}
\includegraphics[width=0.9\textwidth]{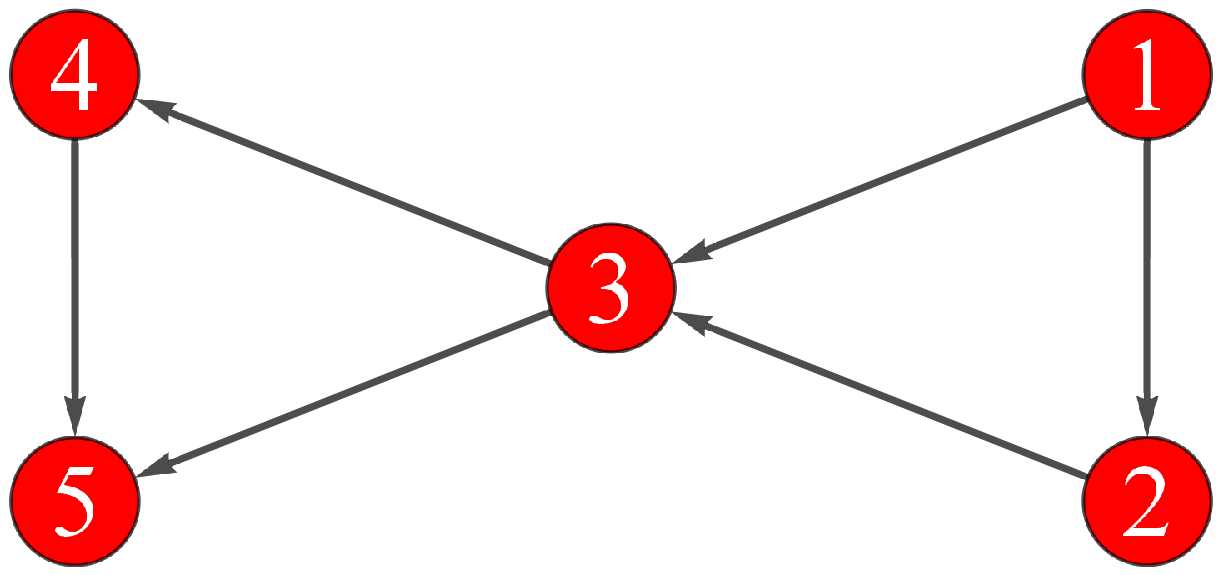}
\end{minipage}\\
Time series&
\begin{minipage}{0.15\textwidth}
\includegraphics[width=0.9\textwidth]{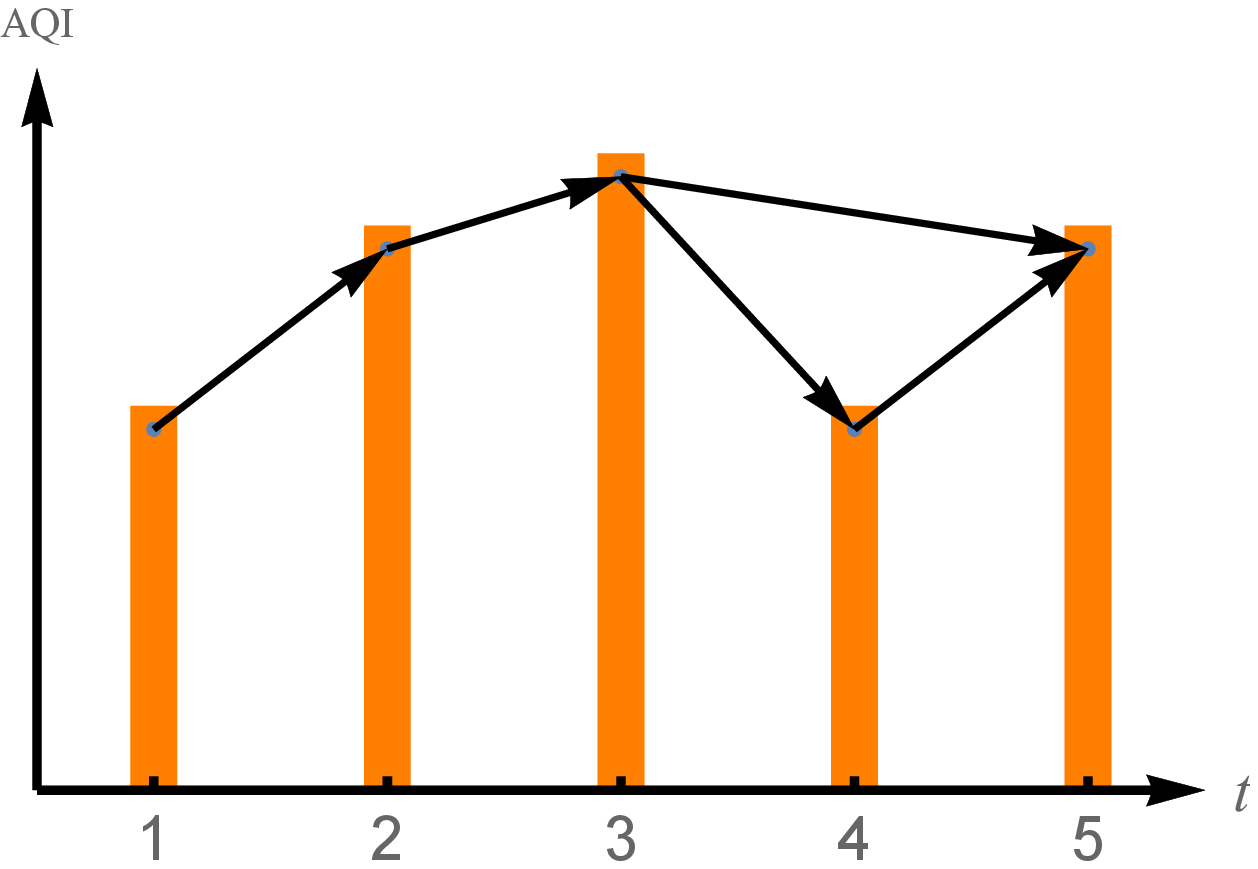}
\end{minipage}&\begin{minipage}{0.15\textwidth}
\includegraphics[width=0.9\textwidth]{motifseries304}
\end{minipage}&\begin{minipage}{0.15\textwidth}
\includegraphics[width=0.9\textwidth]{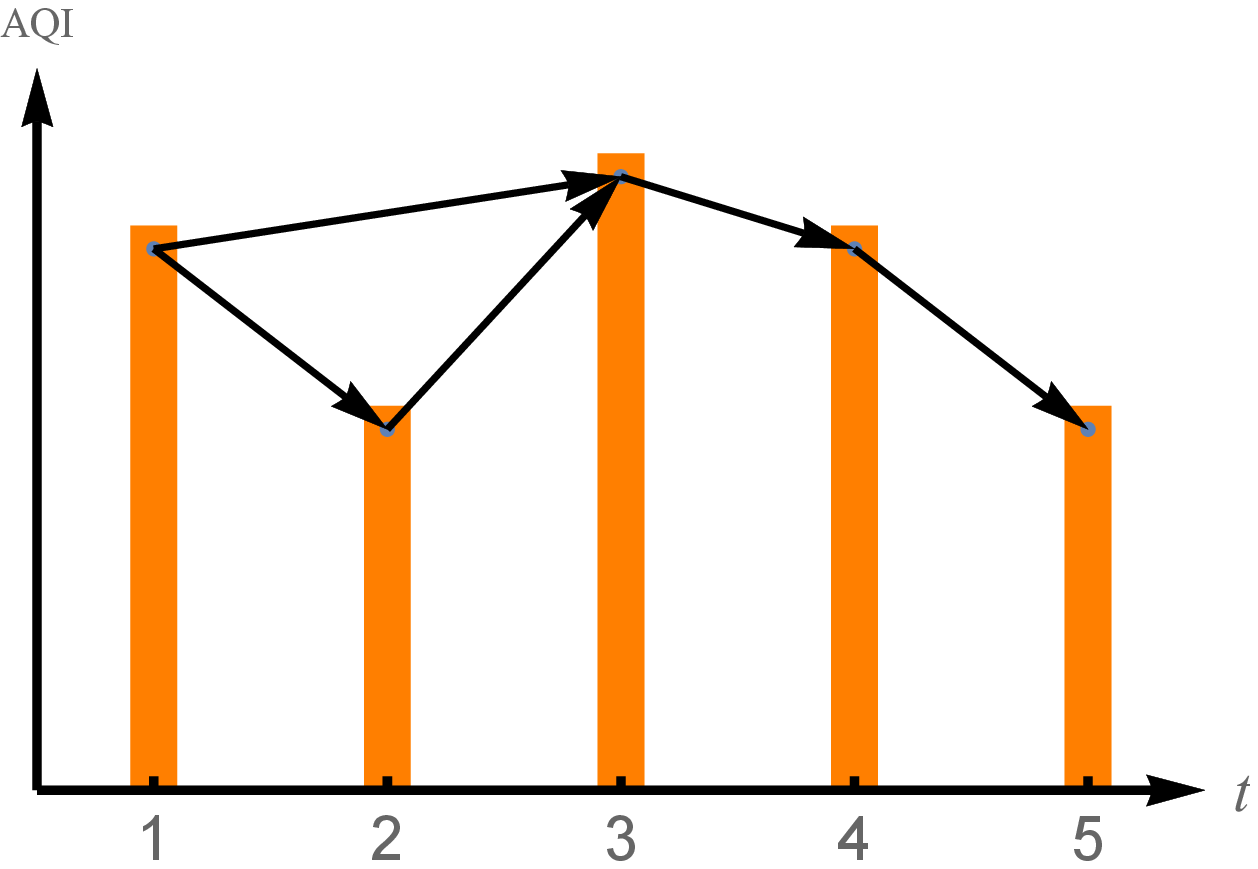}
\end{minipage}&\begin{minipage}{0.15\textwidth}
\includegraphics[width=0.9\textwidth]{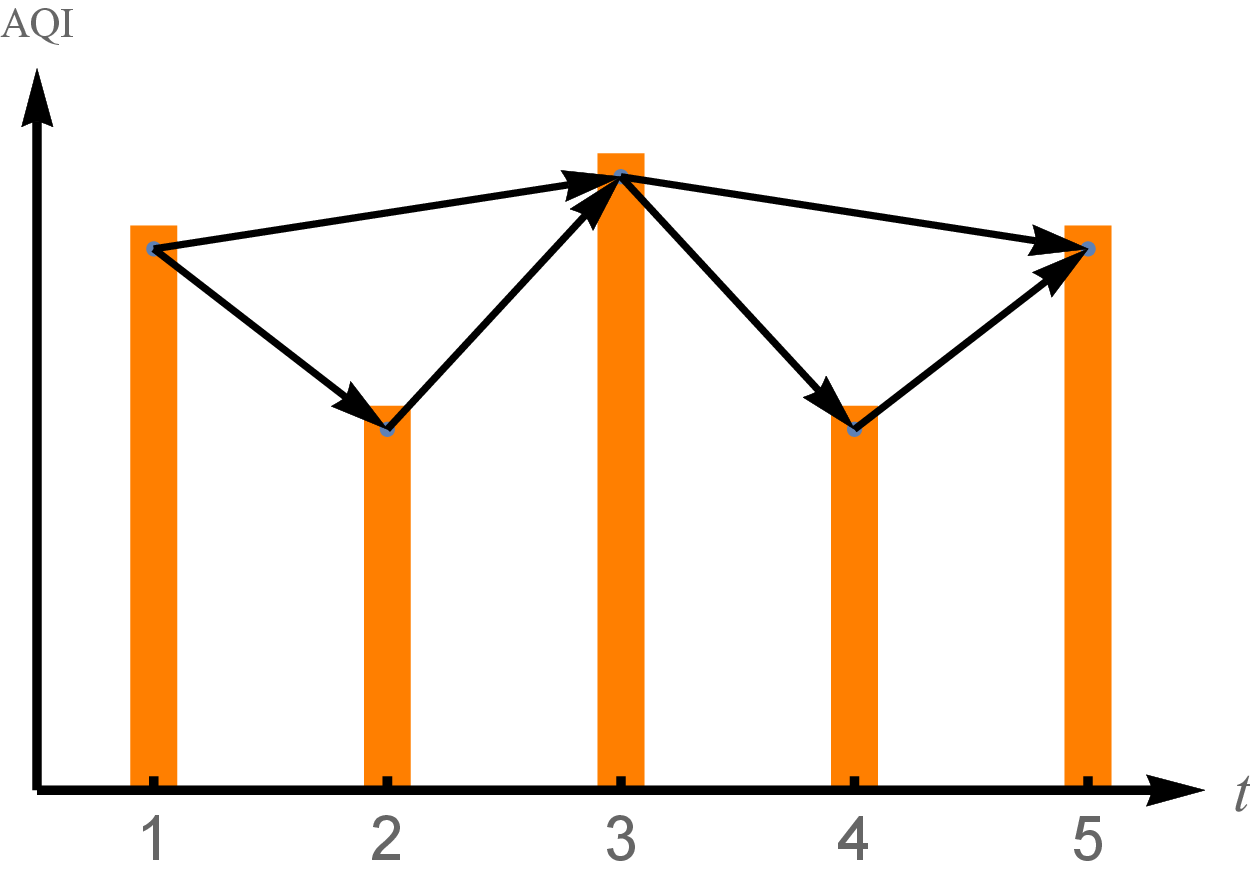}
\end{minipage}\\
\br
\\
\br
Graph label&19&0&311&16\\ 
\hline
Visibility graph&
\begin{minipage}{0.17\textwidth}
\includegraphics[width=0.9\textwidth]{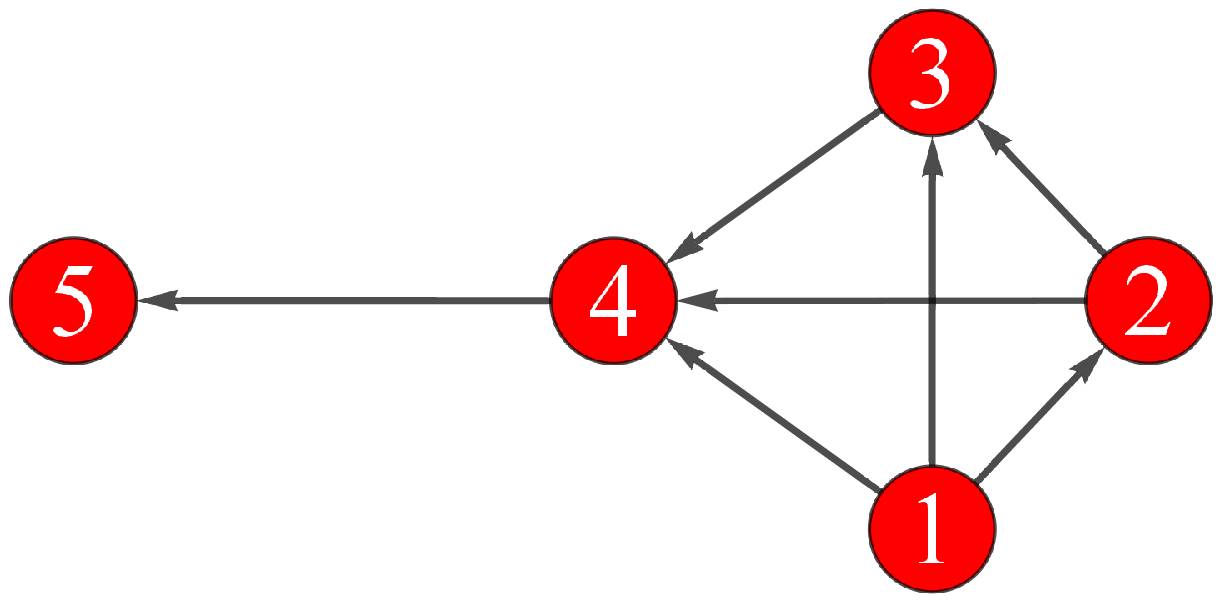}
\end{minipage}&\begin{minipage}{0.15\textwidth}
\includegraphics[width=0.9\textwidth]{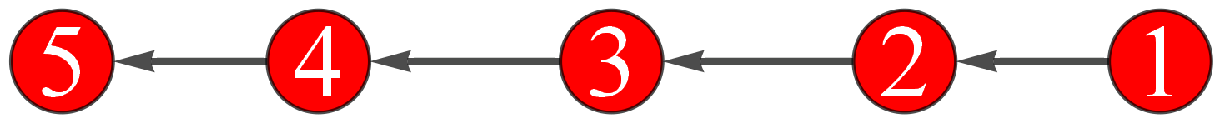}
\end{minipage}&\begin{minipage}{0.15\textwidth}
\includegraphics[width=0.8\textwidth]{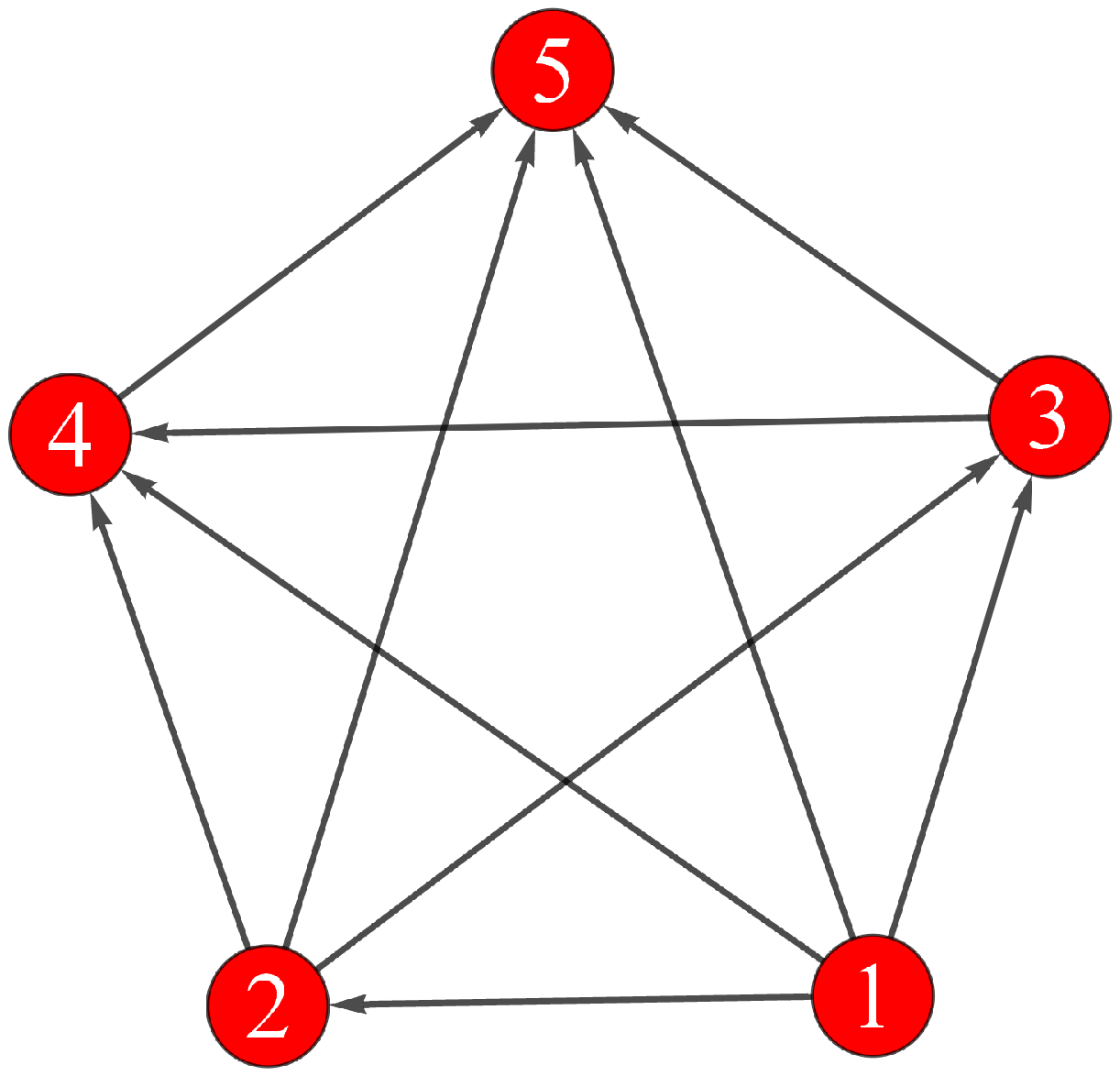}
\end{minipage}&\begin{minipage}{0.15\textwidth}
\includegraphics[width=0.9\textwidth]{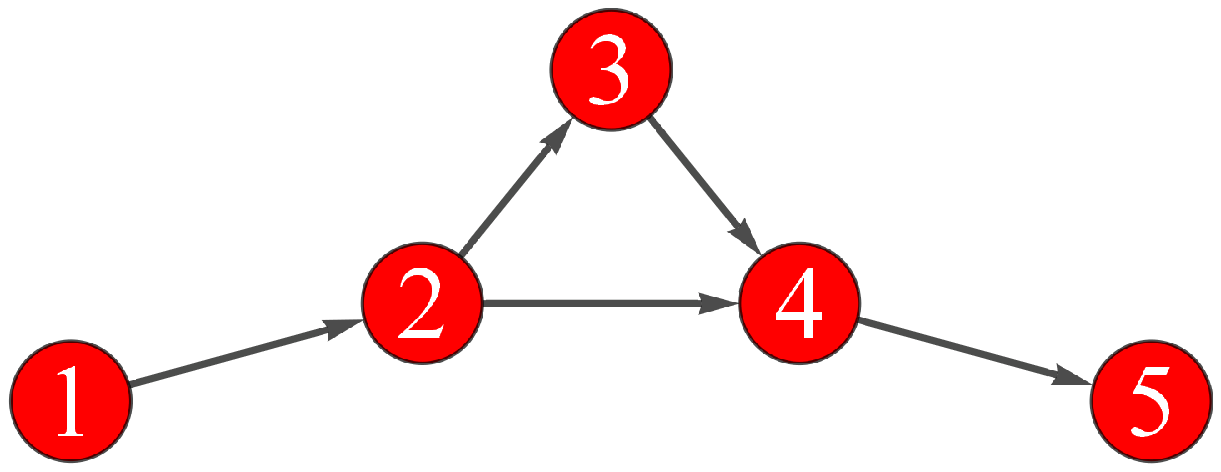}
\end{minipage}\\
Time series&
\begin{minipage}{0.15\textwidth}
\includegraphics[width=0.9\textwidth]{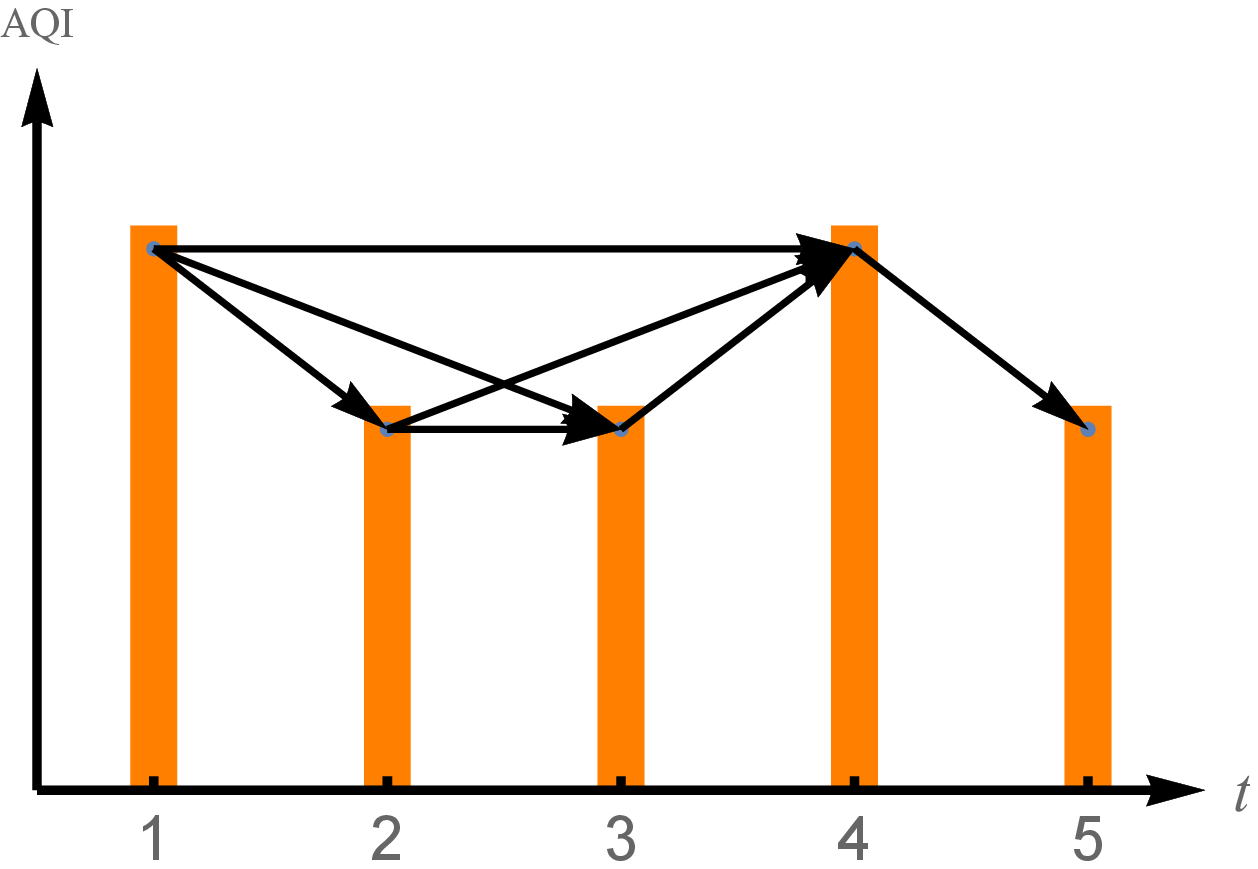}
\end{minipage}&\begin{minipage}{0.15\textwidth}
\includegraphics[width=0.9\textwidth]{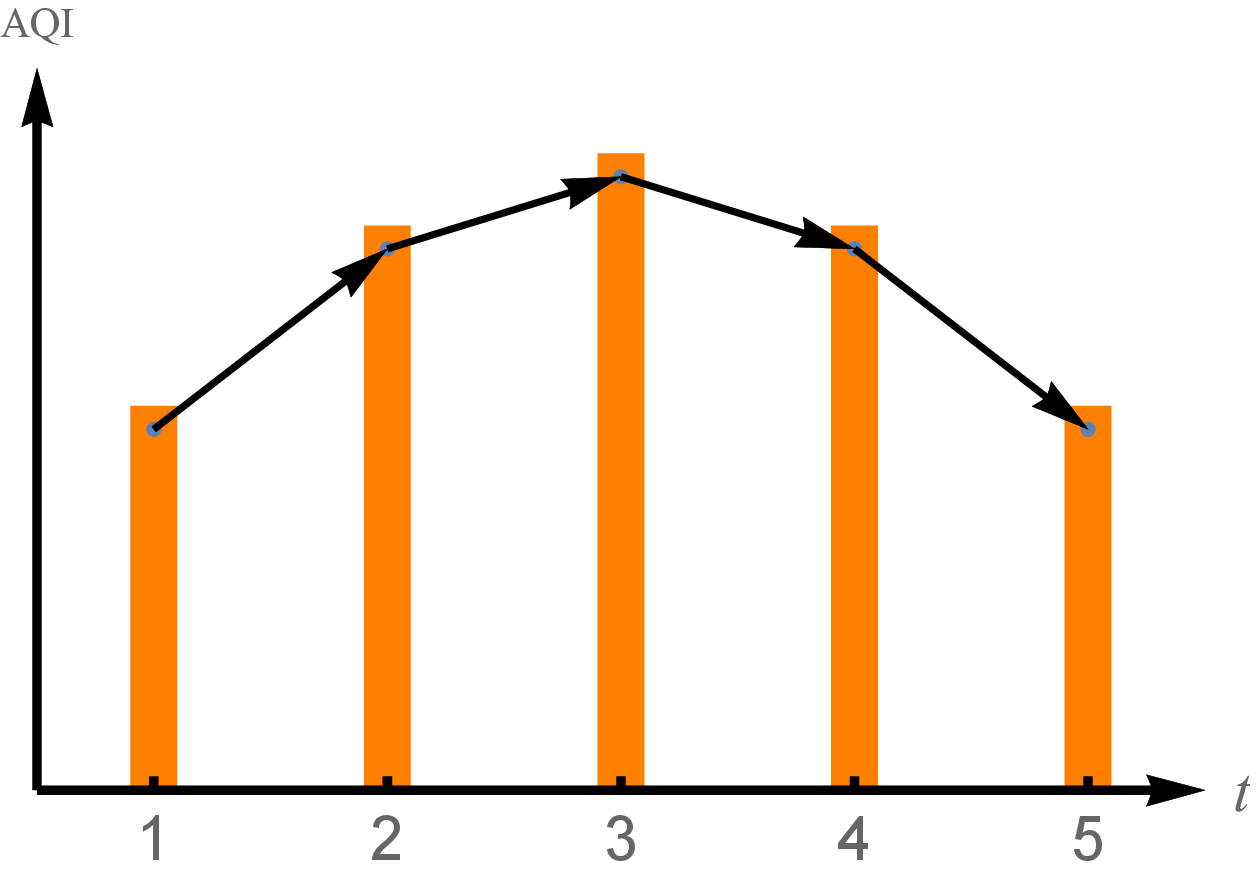}
\end{minipage}&\begin{minipage}{0.15\textwidth}
\includegraphics[width=0.9\textwidth]{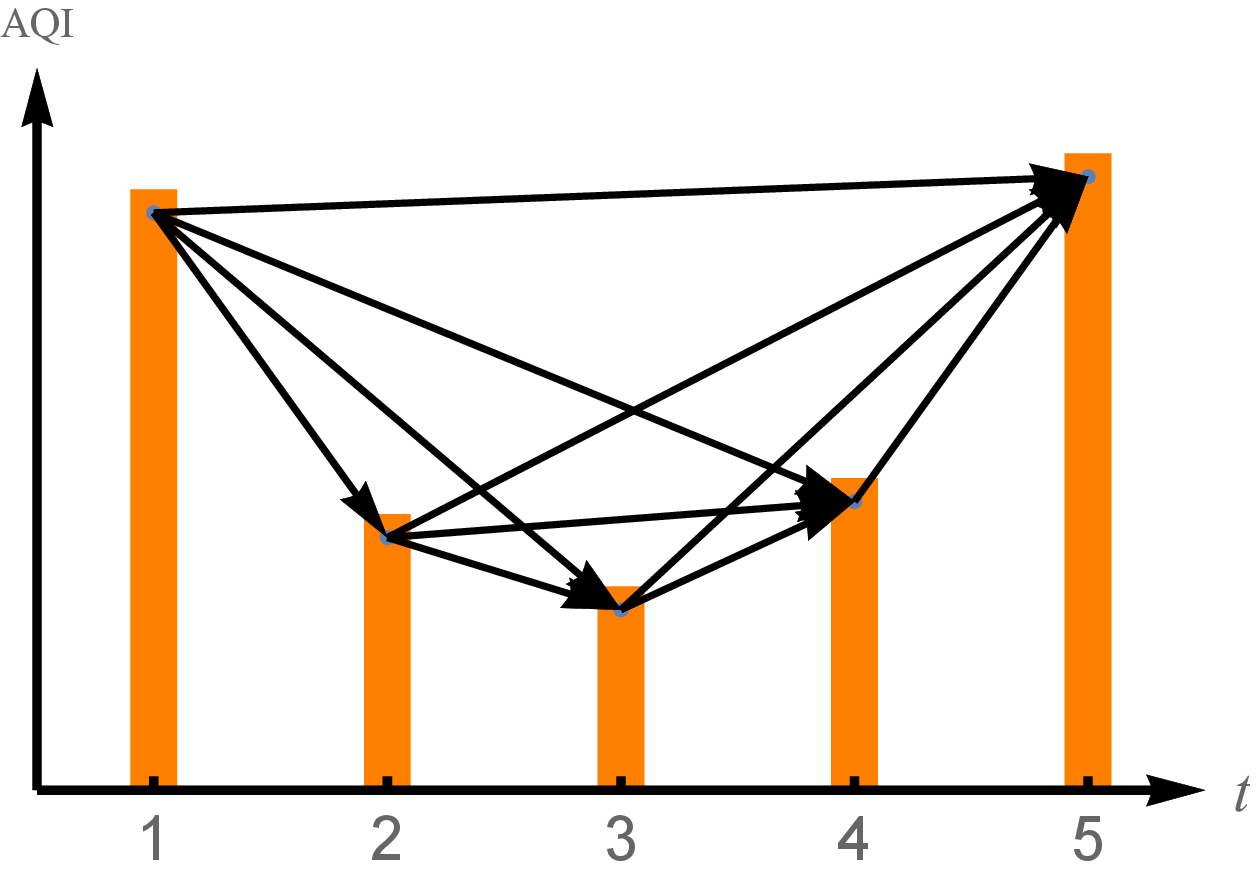}
\end{minipage}&\begin{minipage}{0.15\textwidth}
\includegraphics[width=0.9\textwidth]{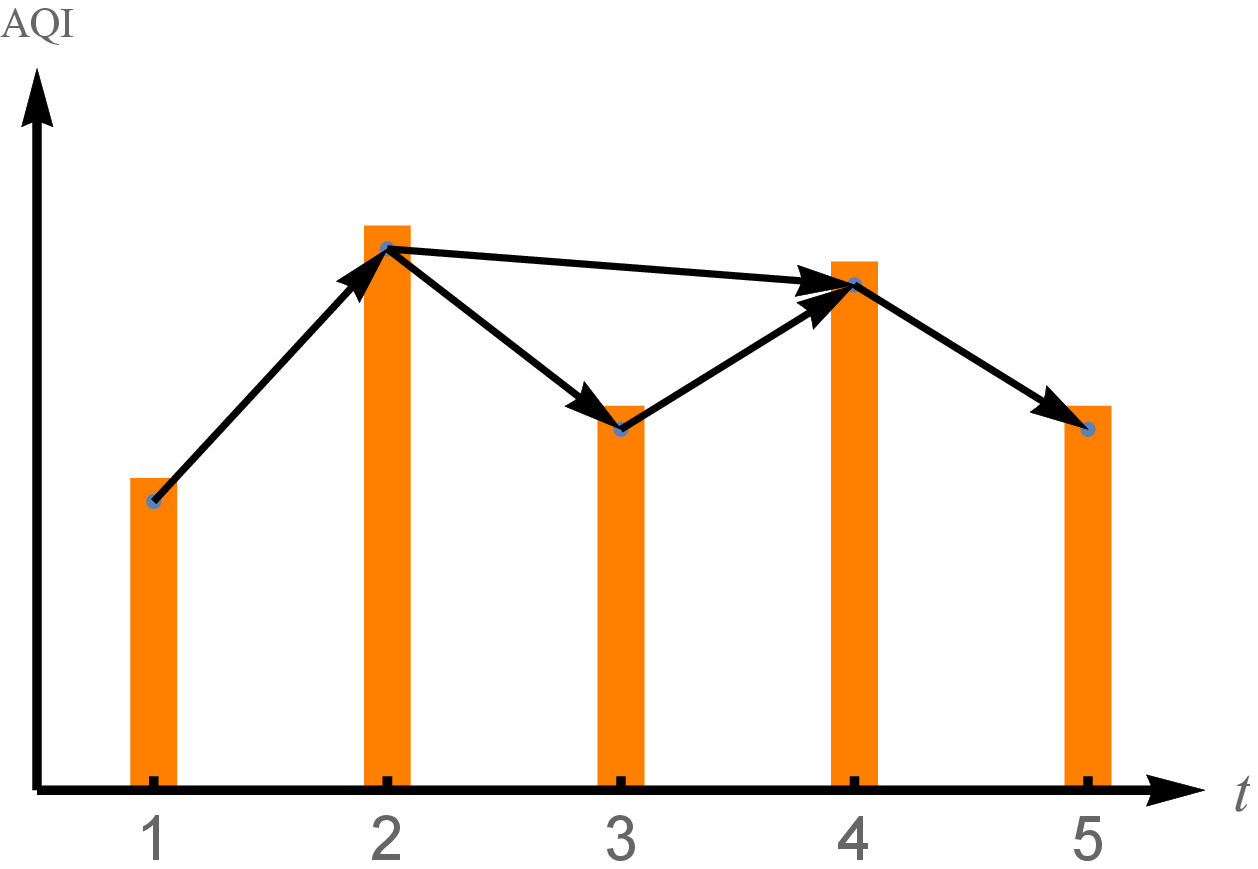}
\end{minipage}\\
\br
\end{tabular}
\end{table}

\subsection{Characteristics of Motifs of the AQI time series}

To investigate the characteristics of motifs of AQI time series, we shuffle the AQI series and construct some new visibility networks, then we compare the frequency of each visibility graph between the two visibility networks. If the degree of an original visibility network is larger than that of the shuffled visibility network, this visibility graph could be  called the ``motif pattern". The different visibility graphs are ranked by the distinction between the original time series and the shuffled one. The top five motifs and corresponding time series of eight communities are shown in Tab.~\ref{table:visibilityg7}. The motifs are not the same with the top frequency visibility graphs (Table.~\ref{table:visibilityg8}), and the motifs $1$ and $2$ of all communities are the same.

\begin{table}
\caption{The top five motifs and corresponding time series of eight communities.}
\label{table:visibilityg7}
\footnotesize
\begin{tabular}{cccccc}
\br
Motif label&1&2&3&4&5\\
 \hline
\multirow{2}*{\shortstack{Southeast\\ China}}&
\begin{minipage}{0.17\textwidth}
\includegraphics[width=0.9\textwidth]{largegraph0}
\end{minipage}&\begin{minipage}{0.13\textwidth}
\includegraphics[width=0.9\textwidth]{largegraph1}
\end{minipage}&\begin{minipage}{0.13\textwidth}
\includegraphics[width=0.9\textwidth]{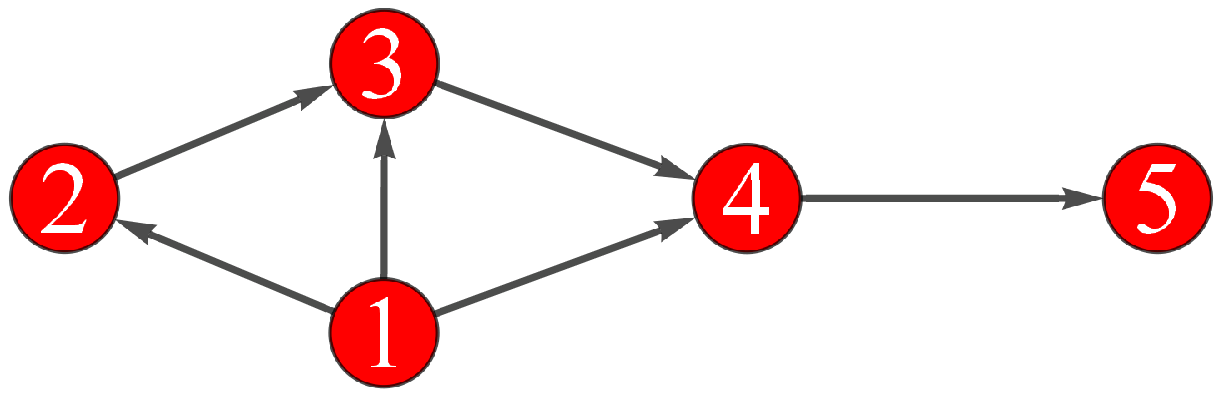}
\end{minipage}&\begin{minipage}{0.13\textwidth}
\includegraphics[width=0.9\textwidth]{largegraph256}
\end{minipage}&\begin{minipage}{0.13\textwidth}
\includegraphics[width=0.8\textwidth]{largegraph311}
\end{minipage}\\
&
\begin{minipage}{0.13\textwidth}
\includegraphics[width=0.9\textwidth]{motifseries0}
\end{minipage}&\begin{minipage}{0.13\textwidth}
\includegraphics[width=0.9\textwidth]{motifseries1}
\end{minipage}&\begin{minipage}{0.13\textwidth}
\includegraphics[width=0.9\textwidth]{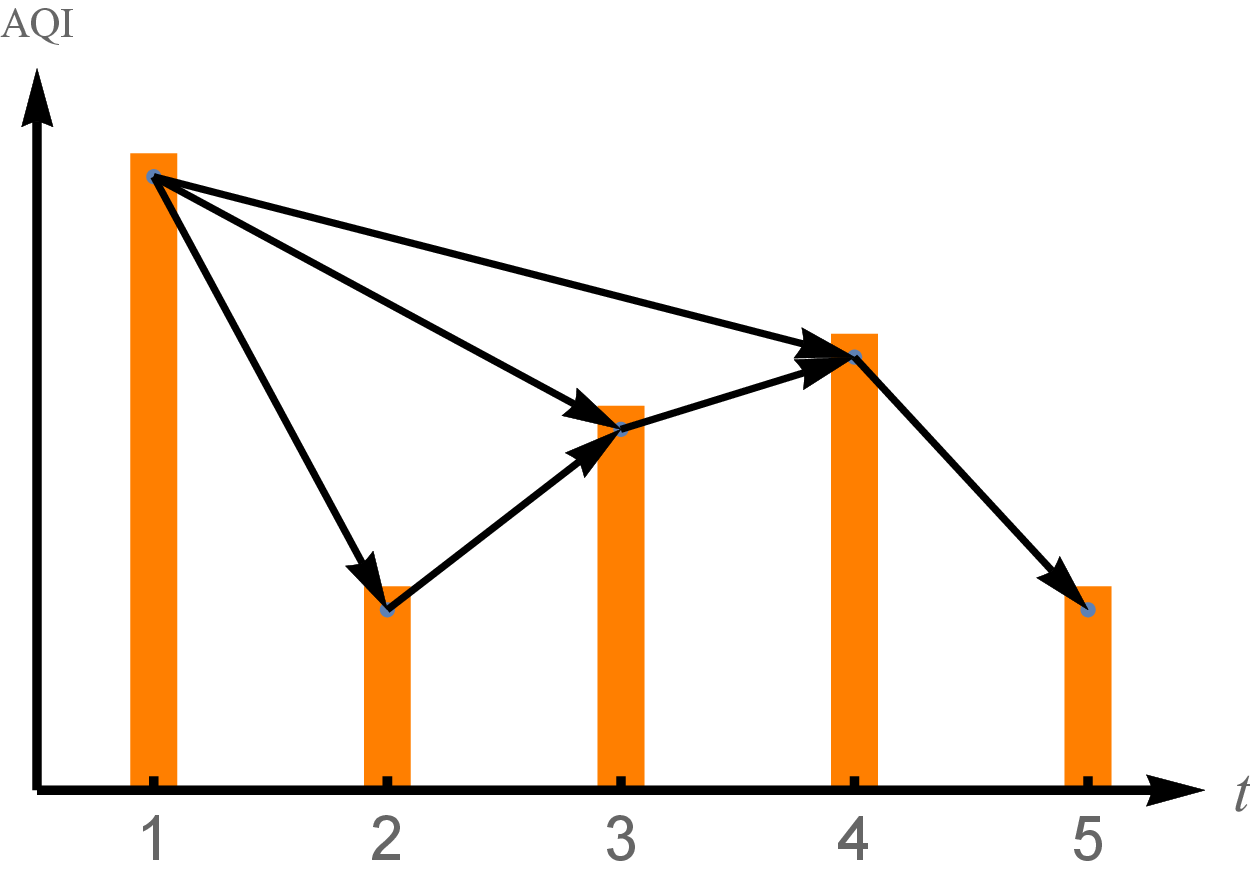}
\end{minipage}&\begin{minipage}{0.13\textwidth}
\includegraphics[width=0.9\textwidth]{motifseries256}
\end{minipage}&\begin{minipage}{0.13\textwidth}
\includegraphics[width=0.9\textwidth]{motifseries311}
\end{minipage}\\

\multirow{2}*{\shortstack{East\\ China}}&
\begin{minipage}{0.13\textwidth}
\includegraphics[width=0.9\textwidth]{largegraph0}
\end{minipage}&\begin{minipage}{0.13\textwidth}
\includegraphics[width=0.9\textwidth]{largegraph1}
\end{minipage}&\begin{minipage}{0.13\textwidth}
\includegraphics[width=0.9\textwidth]{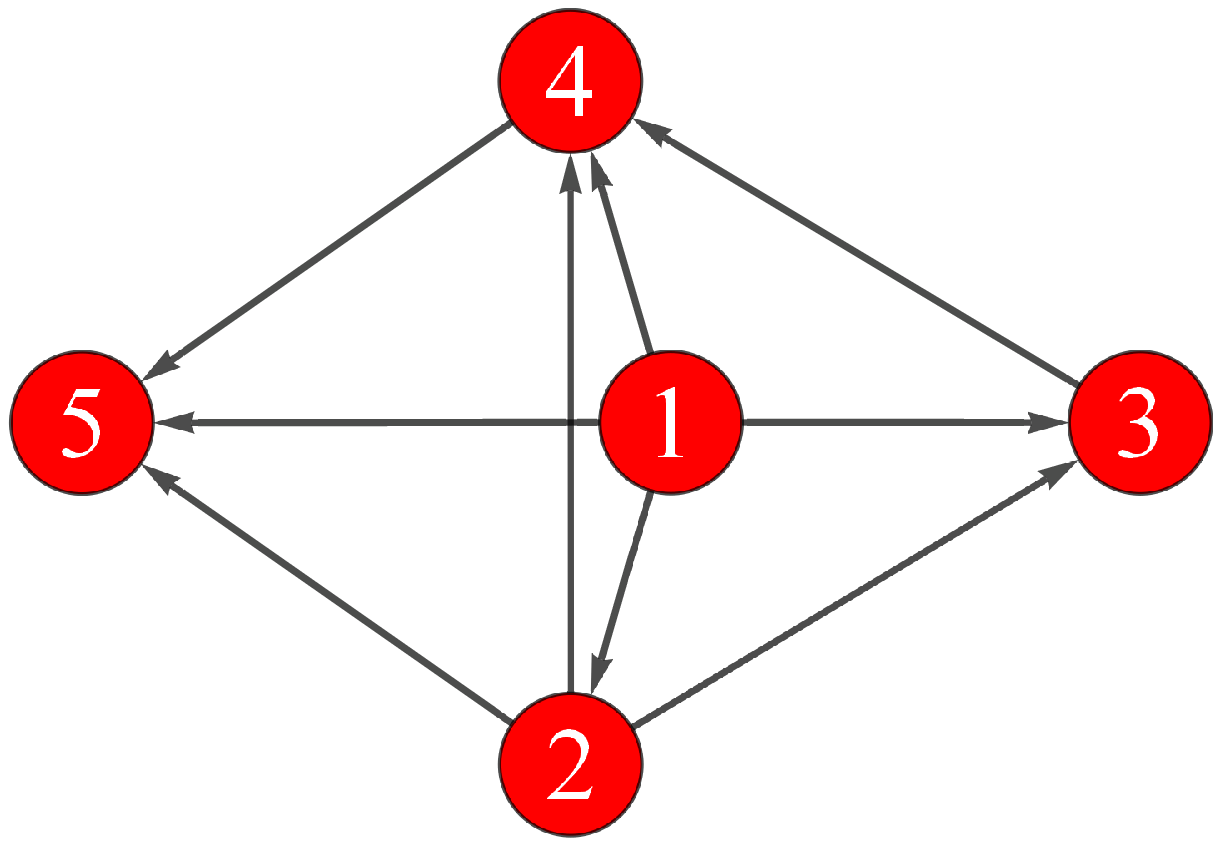}
\end{minipage}&\begin{minipage}{0.13\textwidth}
\includegraphics[width=0.9\textwidth]{largegraph256}
\end{minipage}&\begin{minipage}{0.13\textwidth}
\includegraphics[width=0.8\textwidth]{largegraph311}
\end{minipage}\\
&
\begin{minipage}{0.13\textwidth}
\includegraphics[width=0.9\textwidth]{motifseries0}
\end{minipage}&\begin{minipage}{0.13\textwidth}
\includegraphics[width=0.9\textwidth]{motifseries1}
\end{minipage}&\begin{minipage}{0.13\textwidth}
\includegraphics[width=0.9\textwidth]{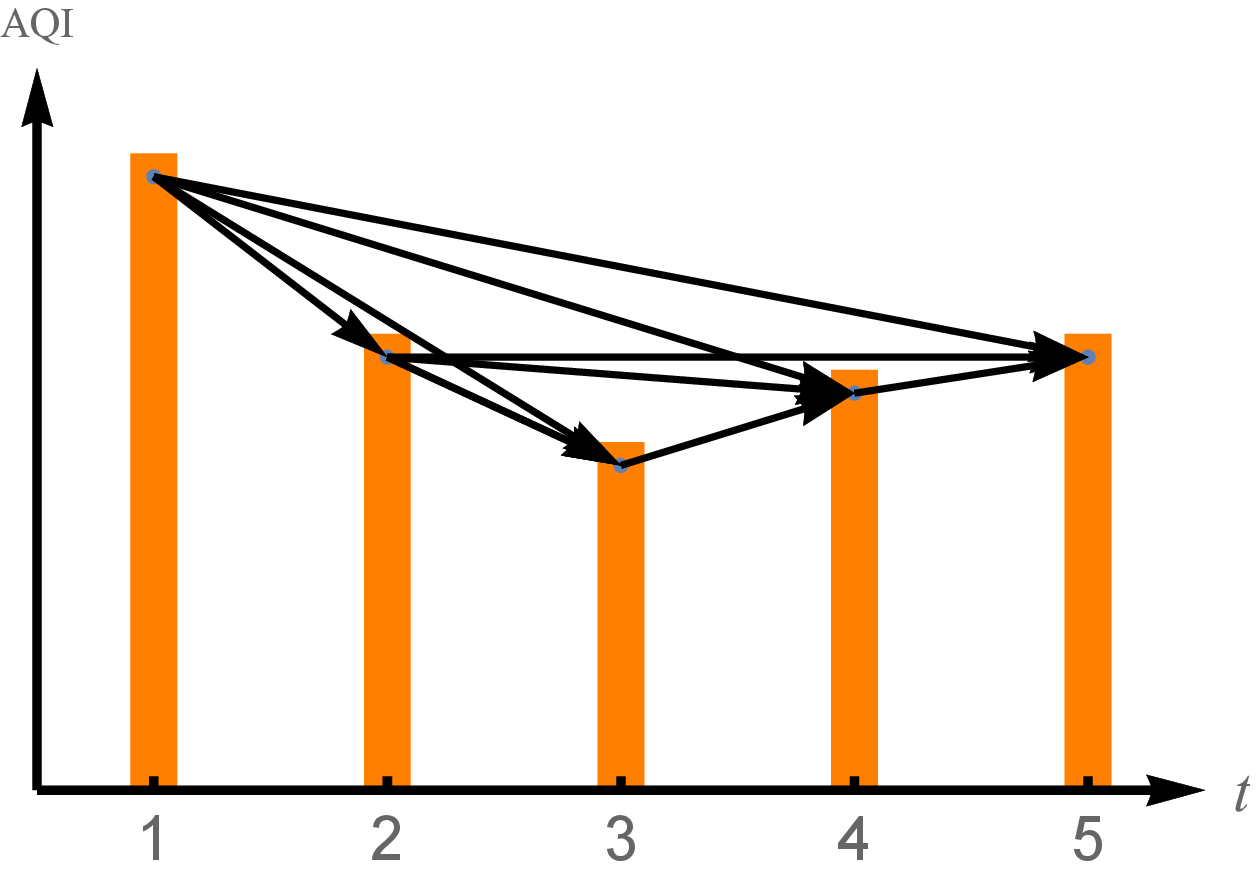}
\end{minipage}&\begin{minipage}{0.13\textwidth}
\includegraphics[width=0.9\textwidth]{motifseries256}
\end{minipage}&\begin{minipage}{0.13\textwidth}
\includegraphics[width=0.9\textwidth]{motifseries311}
\end{minipage}\\

\multirow{2}*{\shortstack{Southwest highland\\ of China}}&
\begin{minipage}{0.13\textwidth}
\includegraphics[width=0.9\textwidth]{largegraph0}
\end{minipage}&\begin{minipage}{0.13\textwidth}
\includegraphics[width=0.9\textwidth]{largegraph1}
\end{minipage}&\begin{minipage}{0.13\textwidth}
\includegraphics[width=0.9\textwidth]{largegraph256}
\end{minipage}&\begin{minipage}{0.13\textwidth}
\includegraphics[width=0.9\textwidth]{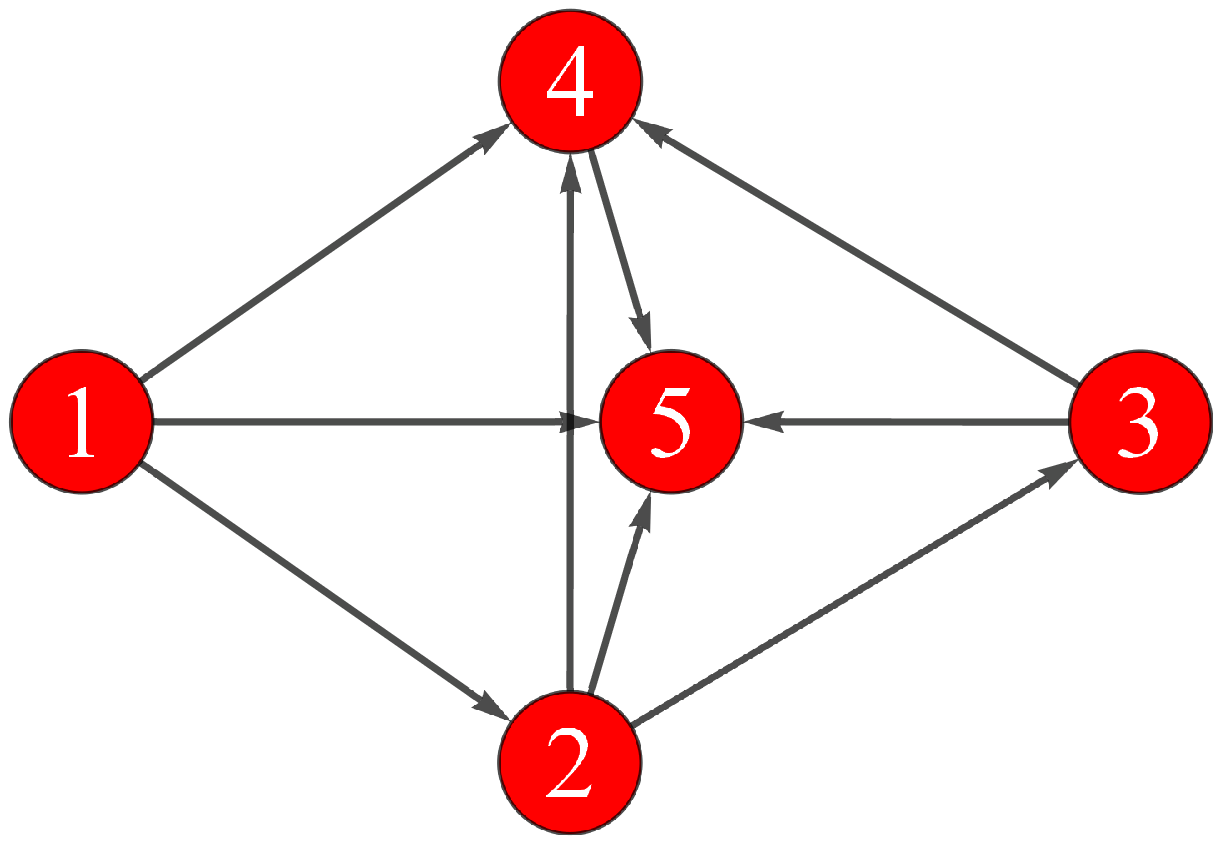}
\end{minipage}&\begin{minipage}{0.13\textwidth}
\includegraphics[width=0.8\textwidth]{largegraph311}
\end{minipage}\\
&
\begin{minipage}{0.13\textwidth}
\includegraphics[width=0.9\textwidth]{motifseries0}
\end{minipage}&\begin{minipage}{0.13\textwidth}
\includegraphics[width=0.9\textwidth]{motifseries1}
\end{minipage}&\begin{minipage}{0.13\textwidth}
\includegraphics[width=0.9\textwidth]{motifseries256}
\end{minipage}&\begin{minipage}{0.13\textwidth}
\includegraphics[width=0.9\textwidth]{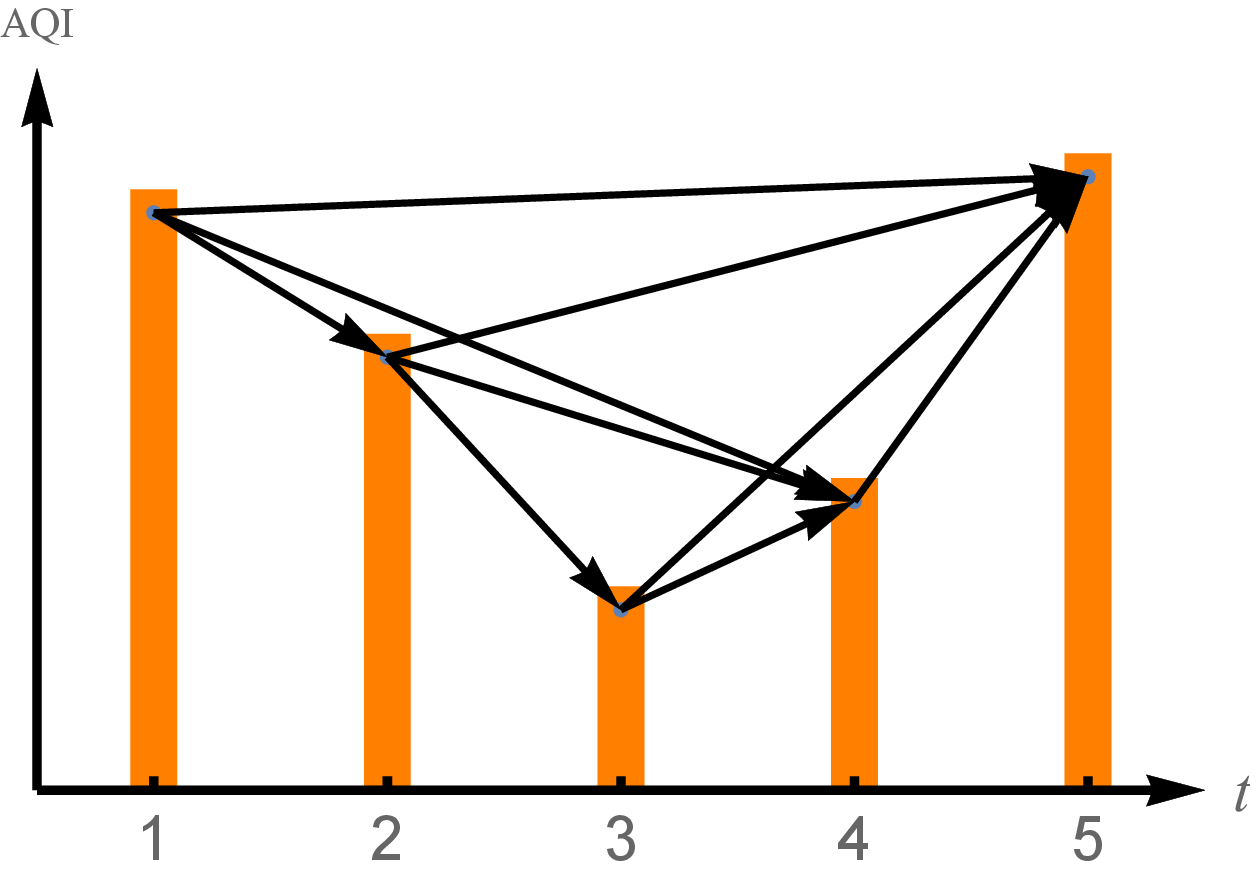}
\end{minipage}&\begin{minipage}{0.13\textwidth}
\includegraphics[width=0.9\textwidth]{motifseries311}
\end{minipage}\\

\multirow{2}*{\shortstack{Northeast \\ China}}&
\begin{minipage}{0.13\textwidth}
\includegraphics[width=0.9\textwidth]{largegraph0}
\end{minipage}&\begin{minipage}{0.13\textwidth}
\includegraphics[width=0.9\textwidth]{largegraph1}
\end{minipage}&\begin{minipage}{0.13\textwidth}
\includegraphics[width=0.9\textwidth]{largegraph55}
\end{minipage}&\begin{minipage}{0.13\textwidth}
\includegraphics[width=0.9\textwidth]{largegraph256}
\end{minipage}&\begin{minipage}{0.13\textwidth}
\includegraphics[width=0.8\textwidth]{largegraph311}
\end{minipage}\\
&
\begin{minipage}{0.13\textwidth}
\includegraphics[width=0.9\textwidth]{motifseries0}
\end{minipage}&\begin{minipage}{0.13\textwidth}
\includegraphics[width=0.9\textwidth]{motifseries1}
\end{minipage}&\begin{minipage}{0.13\textwidth}
\includegraphics[width=0.9\textwidth]{motifseries55}
\end{minipage}&\begin{minipage}{0.13\textwidth}
\includegraphics[width=0.9\textwidth]{motifseries256}
\end{minipage}&\begin{minipage}{0.13\textwidth}
\includegraphics[width=0.9\textwidth]{motifseries311}
\end{minipage}\\

\multirow{2}*{\shortstack{Central-north \\ China}}&
\begin{minipage}{0.13\textwidth}
\includegraphics[width=0.9\textwidth]{largegraph0}
\end{minipage}&\begin{minipage}{0.13\textwidth}
\includegraphics[width=0.9\textwidth]{largegraph1}
\end{minipage}&\begin{minipage}{0.13\textwidth}
\includegraphics[width=0.9\textwidth]{largegraph3}
\end{minipage}&\begin{minipage}{0.13\textwidth}
\includegraphics[width=0.9\textwidth]{largegraph256}
\end{minipage}&\begin{minipage}{0.13\textwidth}
\includegraphics[width=0.9\textwidth]{largegraph310}
\end{minipage}\\
&
\begin{minipage}{0.13\textwidth}
\includegraphics[width=0.9\textwidth]{motifseries0}
\end{minipage}&\begin{minipage}{0.13\textwidth}
\includegraphics[width=0.9\textwidth]{motifseries1}
\end{minipage}&\begin{minipage}{0.13\textwidth}
\includegraphics[width=0.9\textwidth]{motifseries3}
\end{minipage}&\begin{minipage}{0.13\textwidth}
\includegraphics[width=0.9\textwidth]{motifseries256}
\end{minipage}&\begin{minipage}{0.13\textwidth}
\includegraphics[width=0.9\textwidth]{motifseries310}
\end{minipage}\\

\multirow{2}*{\shortstack{Basin of\\ China}}&
\begin{minipage}{0.13\textwidth}
\includegraphics[width=0.9\textwidth]{largegraph0}
\end{minipage}&\begin{minipage}{0.13\textwidth}
\includegraphics[width=0.9\textwidth]{largegraph1}
\end{minipage}&\begin{minipage}{0.13\textwidth}
\includegraphics[width=0.9\textwidth]{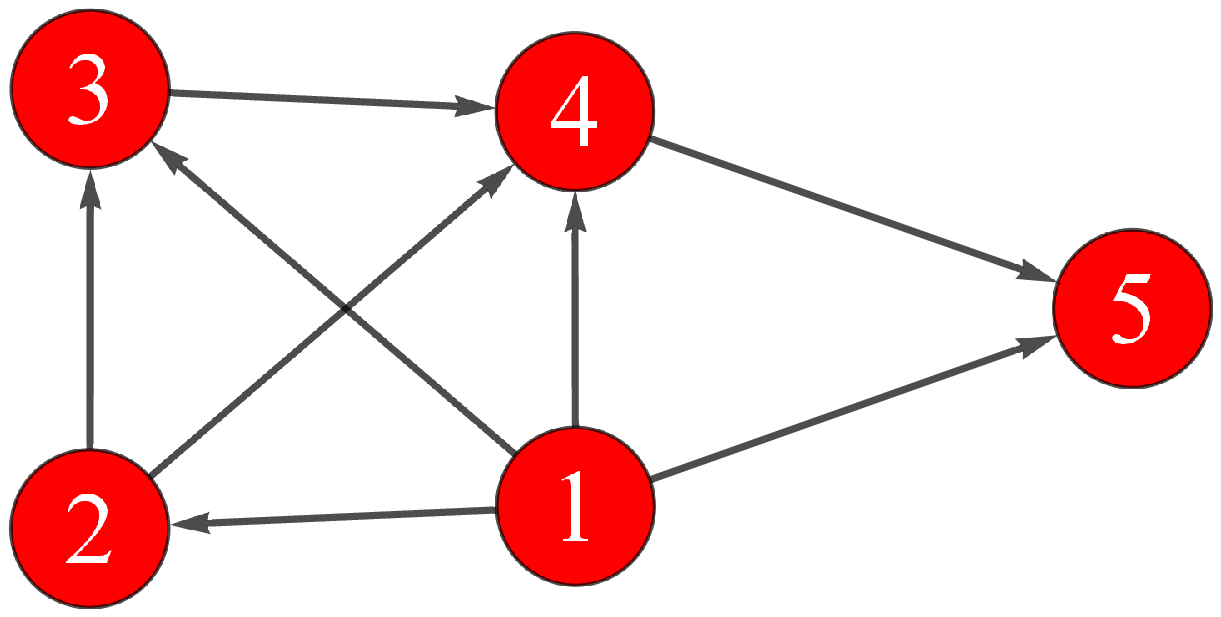}
\end{minipage}&\begin{minipage}{0.13\textwidth}
\includegraphics[width=0.9\textwidth]{largegraph55}
\end{minipage}&\begin{minipage}{0.13\textwidth}
\includegraphics[width=0.9\textwidth]{largegraph256}
\end{minipage}\\
&
\begin{minipage}{0.13\textwidth}
\includegraphics[width=0.9\textwidth]{motifseries0}
\end{minipage}&\begin{minipage}{0.13\textwidth}
\includegraphics[width=0.9\textwidth]{motifseries1}
\end{minipage}&\begin{minipage}{0.13\textwidth}
\includegraphics[width=0.9\textwidth]{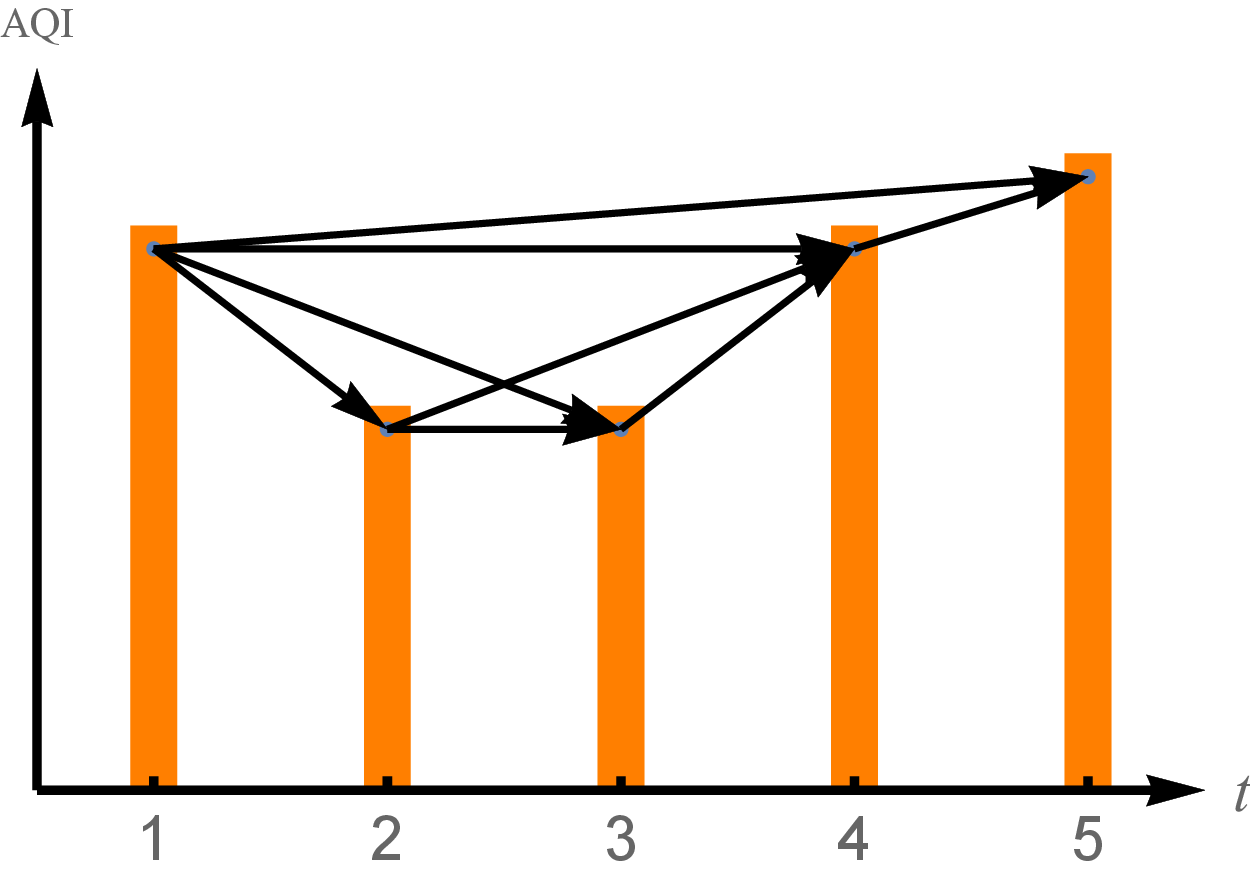}
\end{minipage}&\begin{minipage}{0.13\textwidth}
\includegraphics[width=0.9\textwidth]{motifseries55}
\end{minipage}&\begin{minipage}{0.13\textwidth}
\includegraphics[width=0.9\textwidth]{motifseries256}
\end{minipage}\\

\multirow{2}*{\shortstack{Gobi desert \\of China}}&
\begin{minipage}{0.13\textwidth}
\includegraphics[width=0.9\textwidth]{largegraph0}
\end{minipage}&\begin{minipage}{0.13\textwidth}
\includegraphics[width=0.9\textwidth]{largegraph1}
\end{minipage}&\begin{minipage}{0.13\textwidth}
\includegraphics[width=0.9\textwidth]{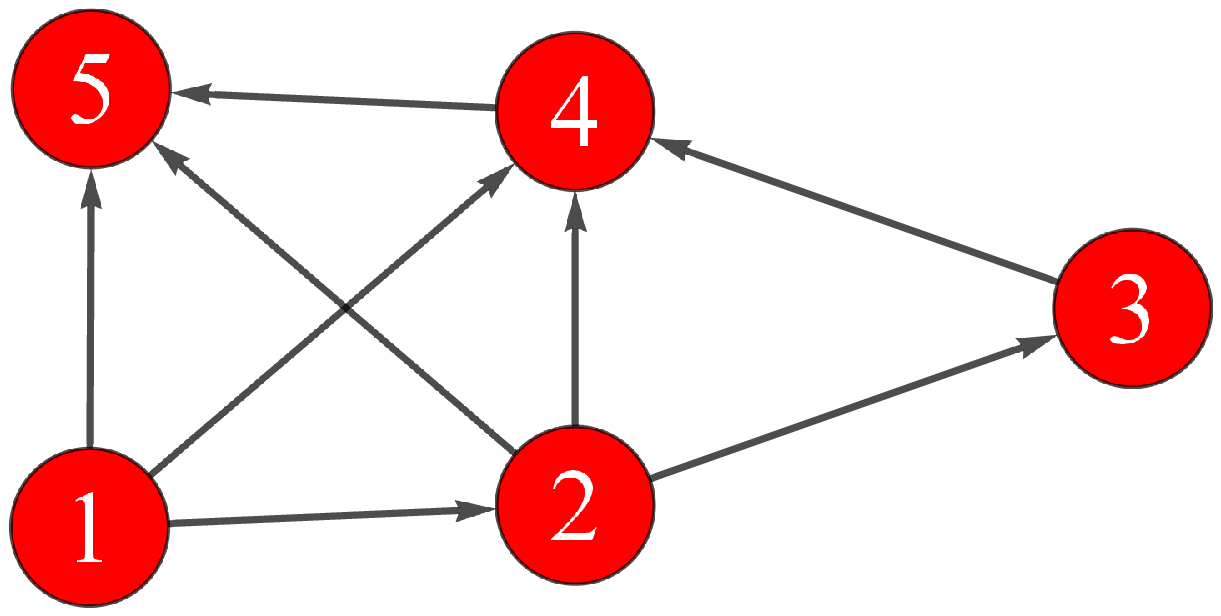}
\end{minipage}&\begin{minipage}{0.13\textwidth}
\includegraphics[width=0.9\textwidth]{largegraph55}
\end{minipage}&\begin{minipage}{0.13\textwidth}
\includegraphics[width=0.9\textwidth]{largegraph310}
\end{minipage}\\
&
\begin{minipage}{0.13\textwidth}
\includegraphics[width=0.9\textwidth]{motifseries0}
\end{minipage}&\begin{minipage}{0.13\textwidth}
\includegraphics[width=0.9\textwidth]{motifseries1}
\end{minipage}&\begin{minipage}{0.13\textwidth}
\includegraphics[width=0.9\textwidth]{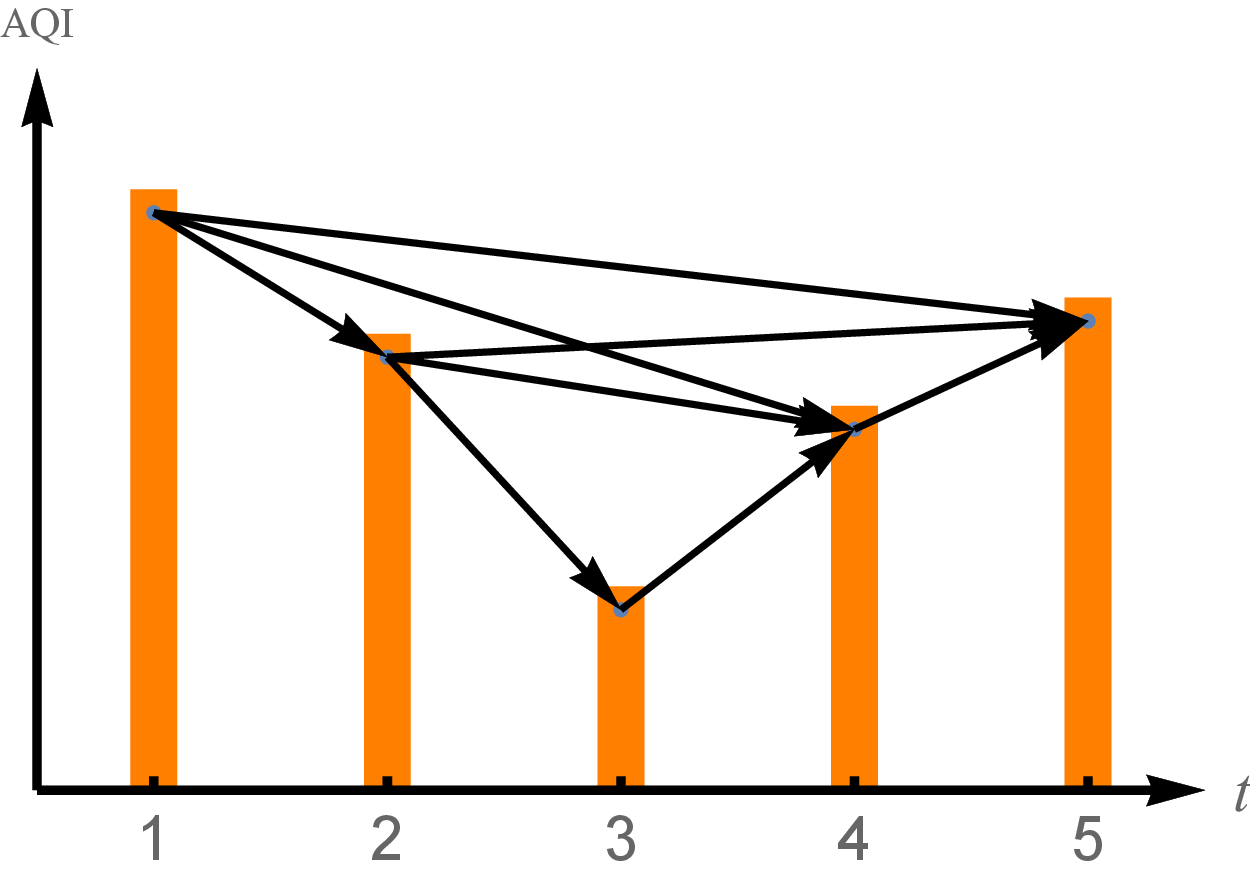}
\end{minipage}&\begin{minipage}{0.13\textwidth}
\includegraphics[width=0.9\textwidth]{motifseries55}
\end{minipage}&\begin{minipage}{0.13\textwidth}
\includegraphics[width=0.9\textwidth]{motifseries310}
\end{minipage}\\

\multirow{2}*{\shortstack{Central-south \\ China}}&
\begin{minipage}{0.13\textwidth}
\includegraphics[width=0.9\textwidth]{largegraph0}
\end{minipage}&\begin{minipage}{0.13\textwidth}
\includegraphics[width=0.9\textwidth]{largegraph1}
\end{minipage}&\begin{minipage}{0.13\textwidth}
\includegraphics[width=0.9\textwidth]{largegraph256}
\end{minipage}&\begin{minipage}{0.13\textwidth}
\includegraphics[width=0.9\textwidth]{largegraph310}
\end{minipage}&\begin{minipage}{0.13\textwidth}
\includegraphics[width=0.8\textwidth]{largegraph311}
\end{minipage}\\
&
\begin{minipage}{0.13\textwidth}
\includegraphics[width=0.9\textwidth]{motifseries0}
\end{minipage}&\begin{minipage}{0.13\textwidth}
\includegraphics[width=0.9\textwidth]{motifseries1}
\end{minipage}&\begin{minipage}{0.13\textwidth}
\includegraphics[width=0.9\textwidth]{motifseries256}
\end{minipage}&\begin{minipage}{0.13\textwidth}
\includegraphics[width=0.9\textwidth]{motifseries310}
\end{minipage}&\begin{minipage}{0.13\textwidth}
\includegraphics[width=0.9\textwidth]{motifseries311}
\end{minipage}\\
\br
\end{tabular}
\end{table}

Then we calculate the Hurst exponents of motifs in eight communities. We implemented the metric \cite{Mutua2016Visibility} of calculating the Hurst exponent of discrete time series with different interval as follows: The top three motifs are chosen to calculate the Hurst exponents of AQI series. Firstly, we denote the time of the chosen motif as $\omega_{k}$, where $k=1,2,...,M$ means the motif is the $k$-th appeared one in the time series. Then we set
\begin{equation}
\Omega^j=(\omega_{j+1}-\omega_j, \omega_{j+2}-\omega_{j+1},...,\omega_{j+n}-\omega_{j+n-1}),
\end{equation}
where $j=1,2,..,M-n$, then
\begin{eqnarray}
\Phi^j(i)&=&\sum_{w=1}^i[\Omega^j(w)-\langle \Omega^j\rangle]\nonumber\\
&=&\omega_{j+1}-\omega_j-\frac{i}{n}(\omega_{j+n}-\omega_j)\quad (i=1,2,...,n.).
\end{eqnarray}

The Hurst exponent can be calculated as
\begin{eqnarray}
\fl R/S(n)=\frac{1}{M-n}\sum_{j-1}^{M-n}\frac{max[\Phi^j(1),\Phi^j(2),...,\Phi^j(n)]-min[\Phi^j(1),\Phi^j(2),...,\Phi^j(n)]}{std(\Omega^j)}.
\end{eqnarray}

\begin{figure}
\centering
\begin{tabular}{c}
\includegraphics[width=0.8\textwidth]{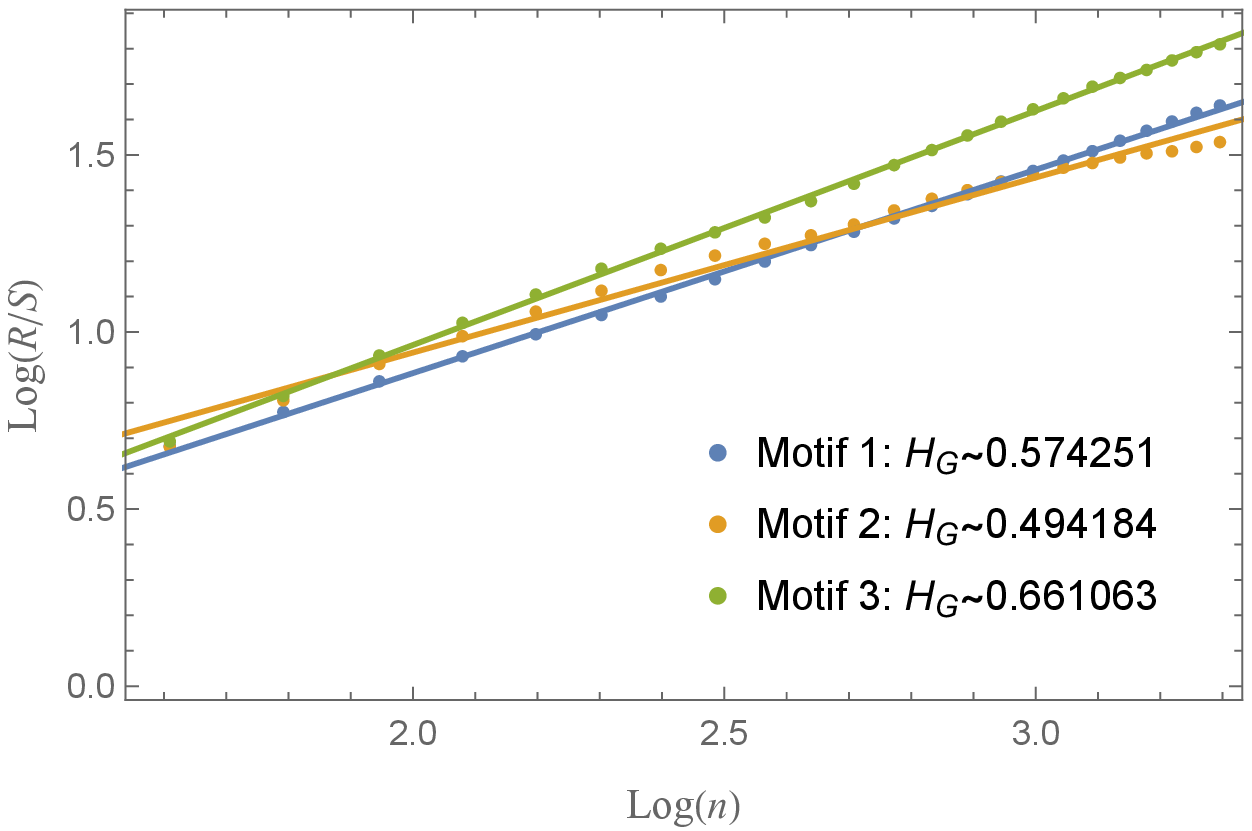}
\end{tabular}
\centering
\caption{(Color online) The relation between $R/S$ and $n$ of motifs, it is plotted in log-log scale. The different colors represent different motifs, and the dotted lines are the calculated results, the fitting results are the real lines.}
\label{fig:RSGraph1}
\end{figure}

In the relation  $R/S(n)\sim n^{H_G}$, $H_G$ is Hurst exponent of the visibility graph pattern. Since the fluctuations of AQI series of single city are large, we average the AQI series of all cities within one community, then we can obtain eight average AQI series. As is the same calculation process of the Hurst exponent, the R/S line becomes not smooth when $n>27$, thus we can calculate $H_G$ by setting $n$ with different values in the range of $(1, 27)$. Otherwise the fourth and the subsequent motifs in each communities have low frequencies, their Hurst exponents could not be calculated, therefore we have only studied the top three motifs. The relation between  $R/S$ and $n$ is presented in Fig.~\ref{fig:RSGraph1} and the $H_{G}$ exponents of all the communities are listed in Table.~\ref{table:Hexponent}. Results show that most of the $H_G$ exponents are larger than $0.5$ or approximately equal to $0.5$. Therefore, the evolutionary patterns have the long-term memory effect and the motifs always appear next if this motif have appear frequently.

\begin{table}
\centering
\caption{\label{table:Hexponent}The $H_G$ exponent of top three motif in $8$ average AQI series, and most $H_G$ exponents are larger than $0.5$ or nearly equals to $0.5$.}
\centering
\begin{tabular}{@{}lcccccccc}

\br
{\bf Community label} &1&2&3&4&5&6&7&8\\ \hline
Motif 1 &0.574& 0.546& 0.512& 0.728& 0.726& 0.639& 0.486& 0.765\\ 
Motif 2 &0.494& 0.623& 0.624& 0.631& 0.619& 0.663& 0.609& 0.503\\ 
Motif 3 &0.661& 0.579& 0.611& 0.735& 0.597& 0.617& 0.531& 0.583\\ 
\br
\end{tabular}

\end{table}

\section*{Conclusion}

We have analyzed the AQI time series from network perspective by using the PMFG method. The correlations of AQI between different cities have been calculated, it is found that there are strong correlations between cities within $100$ km. Eight communities of the AQNC have been found, and we observe that the cities in the same community almost distribute in the same region. We have also calculated the regional Hurst exponent and single city's Hurst exponents, respectively, and found that the AQI time series have strong long-term memory effect. Furthermore, we have used the precipitation, monsoon, and geographical environment to explain the pattern of AQI, regional Hurst exponent, and average Hurst exponent of single cities in every community. Lastly, we have transformed the AQI time series to visibility graphs, and got the motifs of eight communities. The Hurst exponents of motifs have been calculated, and results indicate that the evolutionary patterns of AQI in most communities are have long-term memory.

\section*{Acknowledgements}
We gratefully acknowledge the fruitful discussions with Rongrong Xie, Longfeng Zhao, and Shengfeng Deng. This work was supported in part by National Natural Science Foundation of China (Grant No. 11505071, 11747135, 11905163, 61873104), the Programme of Introducing Talents of Discipline to Universities under Grant No. B08033 the Fundamental Research Funds for the Central Universities (Grant No. KJ02072016-0170, CCNU, CCNU19QN029, CCNU19ZN012), the China Postdoctoral Science Foundation (Grant Nos. 3020501003).

%
%
\section*{References}
\bibliography{iopart-num}

\providecommand{\newblock}{}
\begin{thebibliography}{10}
\expandafter\ifx\csname url\endcsname\relax
  \def\url#1{{\tt #1}}\fi
\expandafter\ifx\csname urlprefix\endcsname\relax\def\urlprefix{URL }\fi
\providecommand{\eprint}[2][]{\url{#2}}

\bibitem{kenneth1998air}
Kenneth W, Warner C~F and Davis W 1998 {\em Third Ediction, Eddison Wesley,
  USA\/}  168--169

\bibitem{Pope1995Particulate}
Pope C~A, Thun M~J, Namboodiri M~M, Dockery D~W, Evans J~S, Speizer F~E and
  Heath C~W 1995 {\em Am J Respir Crit Care Med\/} {\bf 151} 669--674

\bibitem{fakinle2016air}
Fakinle B, Sonibare J, Okedere O, Jimoda L and Ayodele C 2016 {\em Cogent
  Environmental Science\/} {\bf 2} 1208448

\bibitem{Brockwell1989Time}
Brockwell P~J and Davis R~A 1989 {\em Technometrics\/} {\bf 31} 121--121

\bibitem{Schwartz1990Mortality}
Schwartz J and Marcus A 1990 {\em American Journal of Epidemiology\/} {\bf 131}
  185

\bibitem{kim2017ordinal}
Kim S~E 2017 {\em Environmental Modeling \& Assessment\/} {\bf 22} 175--182

\bibitem{Xu2016Spatiotemporal}
Xu L~J, Zhou J~X, Guo Y, Wu T~M, Chen T~T, Zhong Q~J, Yuan D, Chen P~Y and Ou
  C~Q 2016 {\em Air Quality Atmosphere \& Health\/} {\bf 10} 1--9

\bibitem{li2016time}
Li R, Chen Y, Zhao X, Hu Y and Xiao W 2016 {\em Big Data \& Information
  Analytics\/} {\bf 1} 171--183

\bibitem{li2017simple}
Li R, Dong L, Zhang J, Wang X, Wang W~X, Di Z and Stanley H~E 2017 {\em Nature
  Communications\/} {\bf 8} 1841

\bibitem{xu2017clearer}
Xu Y, Li R, Jiang S, Zhang J and Gonz{\'a}lez M~C 2017 Clearer skies in
  beijing--revealing the impacts of traffic on the modeling of air quality
  Report 17-05211 Transportation Research Board 96th Annual Meeting

\bibitem{xu2019unraveling}
Xu Y, Jiang S, Li R, Zhang J, Zhao J, Abbar S and Gonz{\'a}lez M~C 2019 {\em
  Computers, Environment and Urban Systems\/} {\bf 75} 12--21

\bibitem{erdHos1960evolution}
Erd{\H{o}}s P and R{\'e}nyi A 1960 {\em Publ. Math. Inst. Hung. Acad. Sci\/}
  {\bf 5} 17--60

\bibitem{Watts1998Collective}
Watts D~J and Strogatz S~H 1998 {\em Nature\/} {\bf 393} 440

\bibitem{Barabasi1999Emergence}
Barab{\'a}si A~L and Albert R 1999 {\em Science\/} {\bf 286} 509--512

\bibitem{lacasa2008time}
Lacasa L, Luque B, Ballesteros F, Luque J and Nuno J~C 2008 {\em Proceedings of
  the National Academy of Sciences\/} {\bf 105} 4972--4975

\bibitem{zhang2006complex}
Zhang J and Small M 2006 {\em Physical Review Letters\/} {\bf 96} 238701

\bibitem{marwan2009complex}
Marwan N, Donges J~F, Zou Y, Donner R~V and Kurths J 2009 {\em Physics Letters
  A\/} {\bf 373} 4246--4254

\bibitem{yang2008complex}
Yang Y and Yang H 2008 {\em Physica A: Statistical Mechanics and its
  Applications\/} {\bf 387} 1381--1386

\bibitem{bezsudnov2014time}
Bezsudnov I and Snarskii A 2014 {\em Physica A: Statistical Mechanics and its
  Applications\/} {\bf 414} 53--60

\bibitem{gao2017complex}
Gao Z~K, Small M and Kurths J 2017 {\em EPL (Europhysics Letters)\/} {\bf 116}
  50001

\bibitem{fan2016characterizing}
Fan X, Wang L, Xu H, Li S and Tian L 2016 {\em Environmental Science and
  Pollution Research\/} {\bf 23} 3621--3631

\bibitem{carnevale2009neuro}
Carnevale C, Finzi G, Pisoni E and Volta M 2009 {\em Atmospheric Environment\/}
  {\bf 43} 4811--4821

\bibitem{Zhang2018Correlation}
Zhang Y, Chen D, Fan J, Havlin S and Chen X 2018 {\em EPL (Europhysics
  Letters)\/} {\bf 122} 58003

\bibitem{du2019percolation}
Du R, Li J, Dong G, Tian L, Qing T, Fang G and Dong Y 2019 {\em Physica A:
  Statistical Mechanics and its Applications\/}  123312

\bibitem{zhang2018research}
Zhang Y and Na S 2018 {\em Sustainability\/} {\bf 10} 1073

\bibitem{wei2019complex}
Wei Y, Chen L, Qi Y, Wang C, Li F, Wang H and Chen F 2019 {\em
  Sustainability\/} {\bf 11} 3920

\bibitem{tianqihoubao}
\url{http://www.tianqihoubao.com/}

\bibitem{benesty2009pearson}
Benestys J, Chen J, Huang Y and Cohen I 2009 Pearson correlation coefficient
  {\em Noise reduction in speech processing\/} (Springer) pp 1--4

\bibitem{Zhao2016Structure}
Zhao L, Wei L and Xu C 2016 {\em Physics Letters A\/} {\bf 380} 654--666

\bibitem{Kenett2015Network}
Kenett D~Y and Havlin S 2015 {\em Mind \& Society\/} {\bf 14} 155--167

\bibitem{tumminello2005tool}
Tumminello M, Aste T, Di~Matteo T and Mantegna R~N 2005 {\em Proceedings of the
  National Academy of Sciences\/} {\bf 102} 10421--10426

\bibitem{sun2013methods}
Sun P~G and Yang Y 2013 {\em Physica A: Statistical Mechanics and its
  Applications\/} {\bf 392} 1977--1988

\bibitem{china2015ministry}
China M 2015 {\em 2015 Report on the State of the Environment in China\/}

\bibitem{fan2017network}
Fan J, Meng J, Ashkenazy Y, Havlin S and Schellnhuber H~J 2017 {\em Proceedings
  of the National Academy of Sciences\/} {\bf 114} 7543--7548

\bibitem{meng2017percolation}
Meng J, Fan J, Ashkenazy Y and Havlin S 2017 {\em Chaos: An Interdisciplinary
  Journal of Nonlinear Science\/} {\bf 27} 035807

\bibitem{brownlee2017introduction}
Brownlee J 2017 {\em Introduction to time series forecasting with python: how
  to prepare data and develop models to predict the future\/} (Machine Learning
  Mastery)

\bibitem{bloomfield2004fourier}
Bloomfield P 2004 {\em Fourier analysis of time series: an introduction\/}
  (John Wiley \& Sons)

\bibitem{boers2019complex}
Boers N, Goswami B, Rheinwalt A, Bookhagen B, Hoskins B and Kurths J 2019 {\em
  Nature\/} {\bf 566} 373

\bibitem{porter2009communities}
Porter M~A, Onnela J~P and Mucha P~J 2009 {\em Notices of the AMS\/} {\bf 56}
  1082--1097

\bibitem{fan2012secom}
Fan M, Wong K~C, Ryu T, Ravasi T and Gao X 2012 {\em PloS one\/} {\bf 7} e39475

\bibitem{ratti2010redrawing}
Ratti C, Sobolevsky S, Calabrese F, Andris C, Reades J, Martino M, Claxton R
  and Strogatz S~H 2010 {\em PloS one\/} {\bf 5} e14248

\bibitem{tang2012sigma}
Tang C, Li X, Cao L and Zhan J 2012 {\em Journal of theoretical biology\/} {\bf
  306} 1--6

\bibitem{zhang2018statistical}
Zhang W, Deng W and Li W 2018 {\em Physica A: Statistical Mechanics and its
  Applications\/} {\bf 502} 218--227

\bibitem{yang2017overview}
Yang J, Liu Q, Li X and Cui X 2017 {\em Sustainability\/} {\bf 9} 1454

\bibitem{Sun2016Distribution}
Sun G~X, Meharg A~A, Li G, Chen Z, Yang L, Chen S~C and Zhu Y~G 2016 {\em
  Scientific Reports\/} {\bf 6} 20953

\bibitem{landformsmap}
\url{https://zhuanlan.zhihu.com/p/45131479}

\bibitem{Hurst1951Long}
Hurst H~E 1951 {\em Trans. Amer. Soc. Civil Eng.\/} {\bf 116} 770--799

\bibitem{bassingthwaighte1994evaluating}
Bassingthwaighte J~B and Raymond G~M 1994 {\em Annals of biomedical
  engineering\/} {\bf 22} 432--444

\bibitem{kirkby1987hurst}
Kirkby M~J 1987 {\em Earth Surface Processes and Landforms\/} {\bf 12} 57--67

\bibitem{Mutua2016Visibility}
Mutua S, Gu C and Yang H 2016 {\em Chaos An Interdisciplinary Journal of
  Nonlinear Science\/} {\bf 26} 441--449

\bibitem{stephen2015visibility}
Stephen M, Gu C and Yang H 2015 {\em PloS one\/} {\bf 10} e0143015

\end{thebibliography}
\end{document}